\documentclass[aps,pre,twocolumn,pre]{revtex4-1}

\usepackage{setspace}
\usepackage{amsmath}
\usepackage{graphicx}

\usepackage{hyperref}

\begin{document}

%\preprint{}

%\title{Understanding kinetics of ligand dissociation from single binding sites}
%\title{Binding kinetics of ligand dissociation from single binding sites has strong dependence on explicit versus implicit ions}
%\title{Dissociation kinetics of multivalently-bound ligands strongly depends on explicit versus implicit ions}
\title{Effects of electrostatic interactions on ligand dissociation kinetics}
%\title{Explicit versus implicit ions generate quantitatively different ligand-binding site dissociation kinetics}
%Authors
\author{Aykut Erba\c{s}}
\affiliation{Department of Materials Science and Engineering, Department of Molecular Biosciences, and Department of Physics and Astronomy, Northwestern University, Evanston, Illinois 60208, USA}
\author{Monica Olvera de la Cruz}
\affiliation{Department of Materials Science and Engineering, Department of Chemistry, Department of Chemical and Biological Engineering, and Department of Physics and Astronomy, Northwestern University, Evanston, Illinois 60208, USA}
\author{ John F. Marko}
\affiliation{Department of Molecular Biosciences, and Department of Physics and Astronomy, Northwestern University, Evanston, Illinois 60208, USA}


%\author{}
%\thanks{}
%\affiliation{}


\date{\today}

\begin{abstract}

We study unbinding of multivalent cationic ligands  from oppositely charged polymeric binding sites sparsely grafted on a flat neutral substrate. 
Our molecular dynamics (MD) simulations 
are suggested by single-molecule studies of protein-DNA interactions.  
We consider univalent salt concentrations spanning roughly a thousandfold range, together with various concentrations of excess ligands in solution. 
To reveal the ionic effects on  unbinding kinetics of spontaneous and facilitated dissociation mechanisms, we treat electrostatic interactions 
both at a Debye-H\"{u}ckel (DH, or `implicit' ions, i.e., use of
an electrostatic potential with a prescribed decay length) level, as well as by the more precise approach of 
considering all ionic species explicitly in the simulations.
We find that the DH approach systematically overestimates unbinding rates, relative to the calculations where all ion pairs are present explicitly in solution,
although many aspects of the two types of calculation are qualitatively similar.
For facilitated dissociation (FD, acceleration of unbinding by free ligands in solution) 
explicit ion simulations lead to unbinding at lower free ligand concentrations. 
Our simulations predict a variety of FD regimes as a function of free ligand and ion concentrations;
a particularly interesting regime is at intermediate concentrations of ligands where non-electrostatic binding strength controls FD.
We conclude that explicit-ion electrostatic modeling is an essential component to quantitatively tackle problems 
in molecular ligand dissociation, including nucleic-acid-binding proteins.

\end{abstract}

%\keywords{}

\maketitle

\section{Introduction}

Electrostatic and non-electrostatic  interactions between molecular ligands and their binding sites  control many important aspects in biomolecular machinery from gene regulation to molecular recognition. Non-electrostatic contributions (e.g., van der Waals) are usually attributed to interactions between charge-neutral groups, whereas effects due to structural charges (e.g.,  phosphate groups on nucleic acids) and solvated ionic species  are the subjects of electrostatics. Cumulatively, non-electrostatic and electrostatic  interactions determine the lifetime (i.e., inverse of unbinding rate) of a ligand on its binding site.

Experimentally, probing the role of non-electrostatic interactions in the unbinding process is possible, for instance, 
by testing different ligand-receptor pairs by varying one of the partners. However, changing either the ligand or the host molecule inevitably changes the electrostatic interactions as well, since the biomolecules involved, e.g., proteins, are usually amphiphilic and complex structures~\cite{Levy:2007ki,Jones:2003be}. Manipulating solution  strength  is an alternative  way of inferring binding thermodynamics. 
This is mainly because salt ions in solution impose an electrostatic screening length scale (i.e., the Debye length), which can be used as a ruler to probe related kinetic rates (the Debye length defines the volume, in which the electrostatic energy of concentration fluctuations is balanced by the thermal energy). Indeed, many workers have used salt concentration as a tool to study the role of  electrostatic and non-elecrostatic interactions in the dissociation kinetics of biological  ligands~\cite{Record:1976vk, Anderson:1995tg, Anderson:1982wl,Rouzina:1998bw, Manning:1978fy,Privalov:2011kw,VanderMeulen:2008kq}. Extensive studies have shown that the  affinity  of nucleic acid binding proteins~\cite{Privalov:2011kw, Sugimura:2006co,VanderMeulen:2008kq,Koblan:2002hl,Senear:2002fm} and oligocations~\cite{Mascotti:1990tg,Mascotti:1997th,Zhang:1996vn,Datta:2003hl}   decreases with increasing  univalent salt concentration.

%In the solution heterogeneous charges, where smaller and larger charged species are dissolved, smaller species (e.g., univalent salt) is distributed near the larger charges (e.g., proteins or nucleic acids). The distribution of salt near the larger charges depends on the geometry and in biological relevant environment usually it is of cylindrical nature. The unbinding kinetic of two larger species, such as a ligand bound on its bindings site indeed is determined by the concentration of salt as well as structural charges on ligand and binding site.

%Kinetics of  unbinding of a  ligand off its binding site also depends on electrostatic interactions  with surrounding charged species including univalent salt  or other charged ligands competing for the same site.  

Solvated univalent salt ions in solution  can affect the unbinding kinetics of a ligand in various ways.  First, increasing the salt concentration weakens Coulomb interactions between a pair of charges separated by a distance larger than the Debye length. Similarly, the screened Coulomb interactions between  structural charges on the binding site and those on  the ligand can promote dissociation by lowering free energy barriers of binding.  Second, changing the salt concentration can alter the ionic distributions and correlations, particularly near the charged macromolecules~\cite{Misra:1994tg,Misra:1994ii}. Hence, salt-induced competition between various entropic and enthalpic  components can  lead to ligand dissociation. Nevertheless, separation of the salt-related  contributions from other effects is at least approximately possible via extrapolation of dissociation rates to high salt limits~\cite{Record:1976vk, Privalov:2011kw}.

Besides univalent salt ions,  an excess amount of free  ligands in solution can  also facilitate dissociation of a bound ligand by  decreasing the lifetime of the ligand  on the binding site~\cite{Aberg:2016kc, Giuntoli:2015jf,Chen:1jh, Graham:2011cy, Sing:2014dz, Dahlke:2017bn, Aberg:2016kc,Luo:2014ff,Kunzelmann:2010ie,Kamar:2017dd}. The free ligands  can be identical to the initially bound ligand  ~\cite{Kamar:2017dd,Graham:2011cy,Chen:1jh,Gibb:2014kf,Hadizadeh:2016hh,Kim:2012gq}. However, different ligand molecules~\cite{Kamar:2017dd,Kim:2012gq,Pennington:2016es} or even nucleic acid fragments~\cite{Giuntoli:2015jf} can lead to the facilitated  dissociation (FD).
 According to the proposed molecular mechanism for FD, a free ligand binds to an already occupied binding site and destabilizes the complex by forming a ternary complex (i.e., two ligands on the same site)~\cite{Aberg:2016kc,Sing:2014dz,Kamar:2017dd}.  This destabilized complex promotes shorter binding lifetimes for the ligands, while imposing  an upper limit on the unbinding rates.

A recent study on FD has shown that  free ligands with concentration ranges  on the order of  few hundred nano mole can significantly weaken the salt dependence of unbinding rates~\cite{Kamar:2017dd}. The ligand concentrations, at which strong deviations from ligand-free assays were observed, are protein concentrations found in cells. The fact that excess ligand and univalent salt can cause analogous effects on the unbinding kinetics  suggests that  the explicit nature of the ions and structural charges of ligands should be considered carefully.

%Dissociation of  bound ligands from their binding sites are shown to accelerate in the presence of excess amount of  free ligands in solution~\cite{Kamar:2017dd,Chen:1jh, Graham:2011cy,Rebecca, Sing:2014dz, Dahlke:2017bn, Aberg:2016kc}. In this  dissociation mechanism facilitated by initially unbound ligands, instability of originally bound protein is favored by  the ligands competing for the binding site. 

Screened electrostatic potentials of the Debye-Huckel (DH) type have a mean-field nature, and provide a way to account for the decay of electrostatic forces between two charges separated by a distance larger than the Debye length in salt solutions. Although these mean-field  potentials  satisfy the Poisson-Boltzmann equation at long ranges, clearly they cannot  account for the entropy or positional correlations of the screening ions. In addition, micro-dissociation events, whose time scales are likely comparable to the relaxation time of the ionic correlations,  can be smeared out completely in mean field-level treatments. Nevertheless, DH and other mean-field methods are attractive in studying charged biological, as well as synthetic systems, due to their advantage in computational   and  mathematical treatment~\cite{Solis:2000dp,Raspaud:1998hs,Tsai:2016gj,Haas:1999gu,Oberholzer:1999bj,Stigter:2002dy}.

To clearly understand the unbinding kinetics, the role of electrostatic interactions  and the possible interplay between the ionic species and unbound ligands should be investigated.  Given  the variety of methods by which electrostatics can be considered, in this work we aim to answer specific questions including whether the way that electrostatic interactions are treated (i.e., by either modeling all ionic species explicitly or considering their effect via a mean-field DH approach) has any effect on the kinetics of unbinding. 
Furthermore, we inquire whether the electrostatic treatment influences the FD process,  in which  both the non-electrostatic binding energy and ionic contributions are expected to play a major role. Most importantly, we can  compare the results obtained from the different electrostatic treatments to identify various electrostatic contributions to the unbinding mechanism.
We investigate these issues by using Molecular Dynamics (MD) simulations, in which electrostatic interactions are considered carefully to distinguish explicit and mean-field level contributions of charged species. The simulations also allow us to manipulate the non-electrostatic pairwise interactions directly without significantly altering electrostatic interactions. Hence, we can probe  the contributions of non-electrostatic binding strengths and ionic effects on the unbinding kinetics.

In our study, we monitor the unbinding events of single cationic ligands from short polymeric binding sites with opposite charges (Fig.~\ref{fig:fig1}). Many binding sites sparsely placed on a charge-neutral surface  allow us to obtain accurate statistics from uncorrelated binding events. Similar setups are commonly used to experimentally investigate unbinding rates for nucleic acid binding proteins~\cite{Brockman:1999iu}  and are also used in single-molecule studies of protein-DNA interactions~\cite{Kamar:2017dd, Graham:2011cy}. 
In the simulations, we investigate two orders of magnitude  of univalent salt concentration  by considering two approaches that can allow us to distinguish the effect of the ionic component in the kinetic events. In the first approach,  all ions (i.e., salt ions and solvated counterions of the ligands and binding sites) are modeled explicitly, and a pairwise Coulomb potential is calculated between all pairs of ions. We refer to these simulations as ``explicit" throughout this work. In the second approach, the ions are removed from  the simulation boxes, but a DH screened electrostatic potential implicitly takes the  effect of the ions into account. We refer to these simulations  as ``implicit". 

Our MD simulations and data analysis reveal that the implicit DH simulations systematically overestimate the unbinding rates of the ligands, 
and that this discrepancy is more dramatic near physiological salt conditions ($c_\mathrm{s} \approx 100\;$mM). The deviation between explicit and implicit treatments of ionic effects persists in the simulations where we have a ``concentration quench"  (i.e., at the initial step of simulations, all ligands are bound to their binding sites, and there is no ligand in solution), and also in the cases, where the FD effect is triggered by the excess amount of free ligands in solution. The finding that rebinding rates are nearly independent of electrostatic treatment (particularly at physiological salt conditions) indicate that the difference is related to local interactions and is not a simple consequence of the coarse-graining of the ions. Furthermore, aside from investigation of electrostatics, our simulations underline various dissociation regimes, in which the FD process at intermediate ligand concentrations depends on the non-electrostatic binding affinity between the ligands and the binding site. This dependency  disappears at high ligand concentrations.
 
The paper is organized as follows. First, we  describe the general simulation methodology and the two  methods employed  in the simulations to calculate electrostatic interactions. Next, the results obtained for spontaneous dissociation simulations are discussed by considering a multi-step dissociation model. In the next section, we   focus on FD. We discuss our results and possible indications for biological systems together  with future prospects in the Conclusion section.

\section{Methods}
\subsection{Details of MD simulations }
%
%
%The simulation model is designed to mimic the single-molecule studies of ligand--host interactions, in which dissociation of initially bound proteins off their binding sites along a short DNA molecules is monitored in  aqueous medium ~\cite{Kamar:2017dd, others}.

\begin{figure}
\includegraphics[scale=0.4]{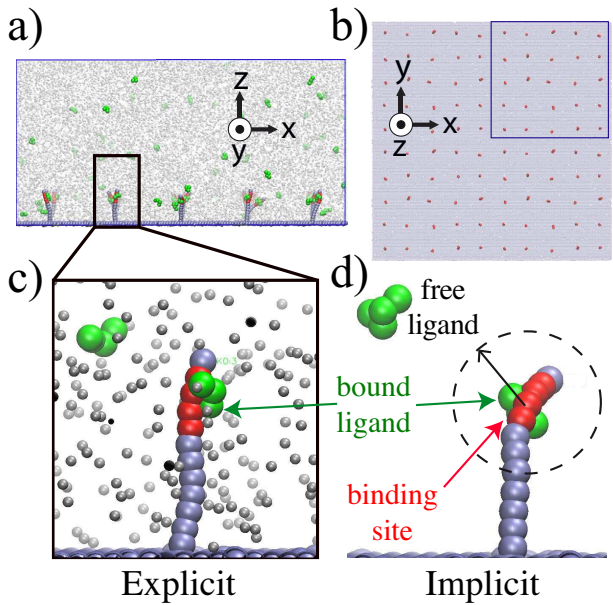}
\caption{a) Side view of a small section of the simulation box with 5 $\times$ 5 binding sites grafted on a flat surface are shown together with free and bound ligands (green beads). Gray beads are univalent ions used only in the explicit simulations, where all  ionic species are treated explicitly. The red beads along the grafted semi-flexible chains attract  bound and free proteins via an attractive short-range potential in addition to the electrostatic interactions.  b) Top view of the actual simulation box with 10 $\times$ 10 binding sites. Ions and ligands are not shown for clarity. c) A single binding site is zoomed in. d) A single binding site used in the implicit-ion simulations, where the electrostatic interactions are calculated via Eq.~\ref{eq:Yukawa}. All beads in the systems except those forming the surface carry respective charges (see the text for details). }
\label{fig:fig1}
\end{figure}

In the simulation model, at least $n_0=10 \times10$ binding sites are sparsely grafted on an inert surface with an inter-site distance of $d=24\;\sigma\;$, where $\sigma$ is the size of  a unit bead (Fig.~\ref{fig:fig1}). The aqueous medium is modeled implicitly as a continuum (see below). The binding sites and ligands  are modeled as coarse-grained “Kremer-Grest (KG)” bead-spring chains~\cite{Kremer:1990bn,Grest:1993uc}. Each binding site is a  linear semi-flexible polymer chain composed of $N = 12$ identical beads.  A  linear chain  of $p=4$  beads is placed onto each binding site to model initially bound ligand molecules (green beads in Fig.~\ref{fig:fig1}). For the FD simulations,  a prescribed concentration of initially free ligands, $c_\mathrm{p}$,  is added at random positions in the simulation boxes in addition to the initially bound ligands. 

The bonding between the adjacent beads of the chains is taken care of by a nonlinear potential with finite extensibility
\begin{equation}
V^{\mbox{\tiny Bond}} (r) = -0.5 k r_0^2 \ln{\left[ 1 - \left(
r/r_0 \right)^2  \right]},
\label{eq:fene}
\end{equation}
where the bond stiffness is $k=30\;\epsilon/\sigma^2$,  the distance between adjacent beads is $r= |\textbf{r}|$, and the maximum bond length is $r_0=1.5\;\sigma$~\cite{Kremer:1990bn}.  The interaction strength $\epsilon$ was measured in the units of thermal energy $k_{\mathrm{B}} T$, where  $k_{\mathrm{B}}$ is the Boltzmann constant, and $T$ is the absolute temperature.

The  steric interactions between all beads are modeled by a
 truncated and shifted Lennard-Jones (LJ) potential, also know as WCA, 
 \begin{equation}
 V^{\mbox{\tiny LJ}}(r) =\left\{ {\begin{array}{*{20}c}
   4\epsilon\left[
(\sigma/ r)^{12} - (\sigma /r)^6 + v_s\right] \;\;\; r \leq r_c \\
   \;\;0~\quad \qquad\qquad\qquad\qquad\qquad\; r > r_c,
\end{array}} \right.
%\theta(r_c-r).
\label{eq:wca}
\end{equation}
where  $r_c$ is the cutoff distance. A cut-off  distance of  $r_c=2^{1/6}\; \sigma$ is used with a shift factor $v_s=1/4$ to obtain good solvent conditions for the interactions between all beads unless otherwise noted  with an interaction strength of 1 $\epsilon$. Four of the $N=12$ beads (the red beads in Fig.~\ref{fig:fig1}) interact  with a strength of 2 $\epsilon$ and with $r_c=2.5\; \sigma$ and $v_s=0$  unless noted otherwise to mimic the specific binding sites for the ligands (e.g., protein specific binding sites along nucleic acid chains).

A harmonic bending potential  is introduced for the  grafted $N=12$ chains to account for the semi-flexible nature of the binding site (e.g.,  DNA)  in the form of 
 \begin{equation}
 V^{\mbox{\tiny Bend}}(\Theta) =k_{ \Theta} \left( \Theta- \Theta_0 \right)^2,
\label{eq:bending}
\end{equation}
where  the potential strength is  $k_{ \Theta}=30~\epsilon / rad^2$,  $ \Theta$ is the angle formed by three adjacent beads, and $ \Theta_0$ is the reference angle. $ \Theta_0=\pi$ for all grafted chains except the beads connecting the chains to the surface. The grafted chains are kept at a right angle with the surface by setting $ \Theta_0=\pi/2$ and $k_{ \Theta} = 90~\epsilon / rad^2$ for the grafted beads.

All  MD simulations are run with LAMMPS MD package~\cite{Plimpton:1995wla} at constant volume $V$ and    $T=1.2~\epsilon/k_{\mathrm{B}}$.  During the relaxation of initial configurations,  each system is simulated for $10^5$ MD steps by keeping the bound ligands on their binding sites by temporarily replacing 1 $\epsilon$ in Eq.~\ref{eq:wca} with 10 $\epsilon$. The data production runs are carried out until a detailed balance is reached for the ligand un/binding events, which is between $10^6 - 10^8$ MD steps. The simulations are run with a time step of $\Delta t=0.005\;\tau$, where the unit time scale in the simulations is $\tau=\sigma \sqrt{m/\epsilon}$ with a monomeric LJ mass of $m=1$. The temperature is kept constant by a Langevin thermostat with a thermostat coefficient $\gamma=0.5\;\tau^{-1}$.
The volume of the total simulation box is set to $L_x \times L_y \times  L_z = 232 \times 232 \times 58\;\sigma^3$. The vertical size of the boxes (i.e.,$L_z = 58\;\sigma$) is higher than the effective effective Gouy-Chapman length of the  surface  due to the grafted charged chains (i.e., $\lambda_{\mathrm{GC}} \approx d^2 / 2 \pi \ell_{\mathrm{B}} N \approx 10\;\sigma$).
% Higher box heights were also tested, but no significant change was observed in unbinding rates.  

\subsection{Calculation of electrostatic interactions in the simulations}
Each bead of the grafted  chains is assigned a unit negative charge, whereas each ligand bead bears a positive unit charge regardless of the method used  for the calculation of the electrostatics. 

In the explicit case, for each charged bead, one oppositely charged counterion bead is added in the simulation box at a random position to obtain  a charge-neutral systems even in the absence of  salt.  Fixed numbers of positively and negatively charged beads are added in the simulation boxes to model salt concentrations in the range of $c_\mathrm{s} \approx 5 - 1000$ mM. The sizes of both counterions and salt monomers  are taken to be $ 0.4\; \sigma$, which is  smaller than the beads forming the binding sites and ligands.  The ionic species interact with each other and with the other components via a shifted 9-6 LJ potential instead of the 12-6 potential given in Eq.~\ref{eq:wca} to account for the effect of (softer) hydration layers. The electrostatic interactions between two charged beads are accounted for by  a pairwise Coulomb potential 
\begin{equation}
V^{\mbox{\tiny Coul}}( r ) =  \pm \epsilon  \ell_{\mathrm{B}} / r  \;\; \text{for}   \;\;r<r_e.
\label{eq:UElec}
\end{equation}
In Eq.~\ref{eq:UElec}, the Bjerrum length $\ell_{\mathrm{B}}$ quantifies the distance at which the Coulomb  energy between two beads of size 1 $\sigma$ at contact is equivalent to   1 $k_{\mathrm{B}}T$.   In aqueous medium, $\ell_{\mathrm{B}}  \approx 0.7$ nm. In the simulations, the dielectric constant is adjusted to obtain $\ell_{\mathrm{B}} =1\; \sigma$. This adjustment corresponds to the charge density  of $\Gamma_{\mathrm{OM}} = 1/\sigma = 1/\ell_{\mathrm{B}}$ as imposed by the Manning condensation effect~\cite{Manning:1969us}.
The electrostatic cut-off  distance in Eq.~\ref{eq:UElec} is    $r_e = 8 - 12\sigma$, above which longer-range electrostatic interactions are set to zero. Particle-Particle-Particle Mesh (PPPM) Ewald solver with an error tolerance $10^{-3}$ was also tested but no significant deviations were observed for the given salt concentrations (see SI Figure S1). Note that the maximum cut-off distance used here is half the distance between two binding sites on the surface.

In the implicit case, all the counterion and salt beads are removed from the simulation boxes, and the electrostatic interactions between the remaining charged species (i.e., ligands and binding-sites)  are calculated via a screened potential

\begin{equation}
V^{\mbox{\tiny Yukawa}}( r ) =  \pm \epsilon  \ell_{\mathrm{B}} / r \exp (-\kappa r),
\label{eq:Yukawa}
\end{equation}
where $\kappa^{-1}$ is the Debye screening length, which is defined as $\kappa^{-1} = k_{\mathrm{B}} T / \sqrt{ 8 \pi  \ell_{\mathrm{B}} c_\mathrm{s}}$ in a  solution of univalent salt. The values of $\kappa^{-1}$  are varied to obtain effective salt concentrations in accord with the explicit-ion simulations. Note that the lowest salt concentration (i.e., $c_\mathrm{s} =5$ mM) considered in this work corresponds to a Debye screening length of $\kappa^{-1}  \approx 4$ nm in real units.
% \approx 0.304 / \sqrt{c_\mathrm{s}}$ nm
%

\subsection{ Extraction of rates}
In the simulations, the number of ligands remaining bound on their binding sites, $n(t)$, is monitored as a function of the simulation time, $t$. If any bound ligand diffuses out of a spherical volume with a radius $R_c \approx  4 \sigma$, centered around the center of mass of the binding site (red beads in Fig.~\ref{fig:fig1}), the ligand is tagged as unbound. If the ligand returns to the binding site, it is not counted as bound. To determine the off-rates both in the absence and presence of unbound ligands, the survival fraction data is fit by a single exponential 
\begin{equation}
n(t)=n_0 \exp(-k_{\mbox{\tiny off}} t),
\label{eq:fraction_of_sites}
\end{equation}
where $k_{\mbox{\tiny off}}$ is defined as the inverse of the lifetime of the bound ligands  on their  binding sites (i.e., $k_{\mbox{\tiny off}} \equiv 1/\tau_{\mbox{\tiny off}}$).  Error bars are calculated by averaging the results of multiple runs. In addition, $n_0$ in Eq.~\ref{eq:fraction_of_sites} is released as a fit parameter to enrich error statistics. In the fitting procedures, a weight function inversely proportional to the square of the data point is used. Error bars are not shown if they are smaller than the size of the corresponding data point. VMD is used for the visualizations~\cite{vmd}. 

%%%%%%%%%%%%%%%%%%%%%%%%%%%%%%%%%%%%%%%%%%%%%%%%%%%%%%%%%
%%%%%%%%%%%%%%%%%%%%%%%%%%%%%%%%%%%%%%%%%%%%%%%%%%%%%%%%%
\section{Results}
%%%%%%%%%%
\subsection{Spontaneous dissociation}
\begin{figure}
\centering
\includegraphics[width=9cm,viewport= 0 0 550 800, clip]{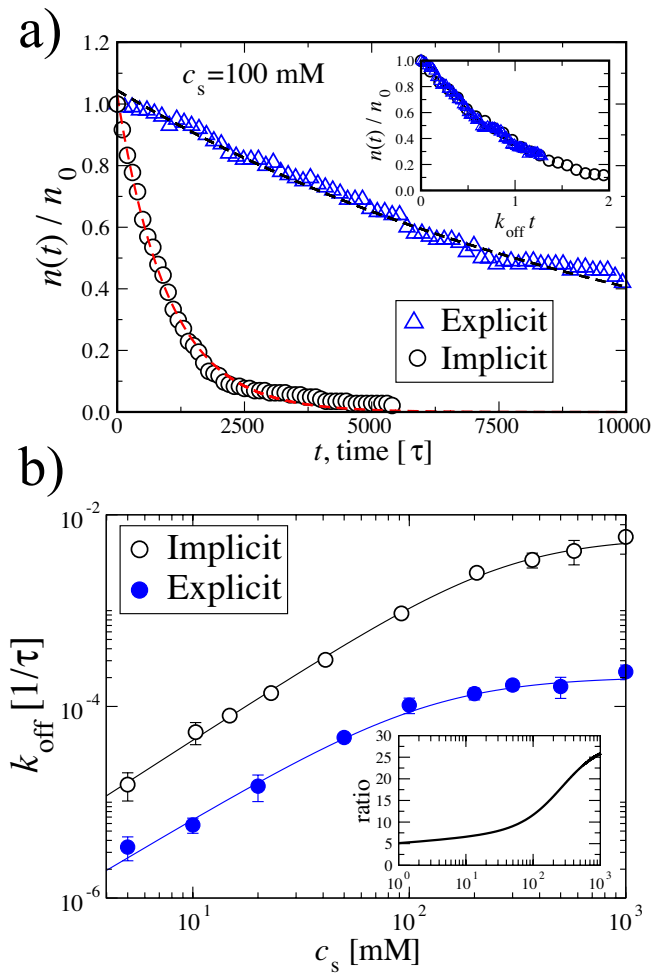}
\caption{a) The survival fractions of bound ligands are shown as a function of the simulation time for the explicit and implicit electrostatics cases. At $t=0$, all sites are occupied (i.e., $n(t=0)=n_0$). The dashed curves are exponential fits to Eq.~\ref{eq:fraction_of_sites}. Inset shows the same time traces but the time axes are rescaled by corresponding values of the off-rates. b) The off-rates obtained from exponential fits are shown as a function salt concentration for both explicit and implicit cases. The curves are obtained via Eq.~\ref{eq:fit_rates}. The inset is the ratio of blue and black curves to highlight the difference with varying salt concentrations.}
\label{fig:fig2}
\end{figure}
\subsubsection{Off-rates for explicit salt are significantly less than those for implicit salt}
We first focus on the case, in which initially all  binding sites  are occupied by single  ligands, and there is no free ligand in solution. This setup  corresponds to  a “concentration quench”, from which an initially high concentration of bound ligands desorbs from the binding sites to an initially zero-concentration bulk state.  Similar setups have commonly been  used in  experiments  studying unbinding rates for protein-DNA interactions~\cite{Kamar:2017dd,Graham:2011cy,Brockman:1999iu}.

In Fig.~\ref{fig:fig2}a, we show typical time traces for the normalized fraction, $n(t)/n_0$,  of bound ligands that are bound at time $t$, as a function of simulation time for both explicit (triangles in Fig.~\ref{fig:fig2}a) and implicit (open circles) cases at a salt concentration of $c_\mathrm{s} \approx 100\;$mM. As  time progresses,   chemical equilibrium between bound and unbound states is established, and all of the initially bound ligands leave their binding sites. For both explicit and implicit cases, the relaxation of the concentration quenches  can be described well by single-exponential functions (dashed curves  in Fig.~\ref{fig:fig2}a). 
However, the off-rate, $k_{\mathrm{off}}$, (i.e., the inverse dwelling  time of the ligand on its binding site) obtained from the exponential fits for the explicit case is one order of magnitude slower than that for the implicit case. This indeed shows that the ligands remain  on their binding sites for less time in the implicit simulations. 
Once the time axes  are rescaled by the corresponding  values of the off-rates, the data sets for the implicit and explicit cases are nearly indistinguishable (see the inset of Fig.~\ref{fig:fig2}a). This indeed confirms that both processes can be described by single-exponential decays independent of how we treat electrostatics.

To compare dependencies of the off-rates on univalent salt concentration (or on the Debye length), we performed similar titration simulations by varying the range of excess salt concentration between $c_\mathrm{s} \approx 5-1000\;$mM. Consistently, the simulations, in which  we consider electrostatic interactions implicitly (open circles Fig.~\ref{fig:fig2}b) systematically exhibit higher unbinding rates  compared to the explicit case (closed circles Fig.~\ref{fig:fig2}b) for the entire range of the salt concentration. 
However, in both cases, the off-rates have similar qualitative dependencies on the excess salt; at low salt, the off-rate increases gradually up to $c_\mathrm{s} \approx 100\;$mM, above which a saturation regime appears. In these saturation regimes, the salt has 
little or no effect on the off-rates. It is noteworthy that beyond $c_\mathrm{s} > 100\;$, the  electrostatic screening length of the solution is less than the size of a unit bead (i.e., $\kappa^{-1} <1\; \sigma$). The plateau values of $k_{\mathrm{off}}$ for the two cases are significantly different as seen in Fig.~\ref{fig:fig2}b. %This hints the effect of explicit nature of ionic atmosphere near the binding sites.

To make a more systematic comparison  between the off-rates obtained from the implicit and explicit cases and to quantify the observed regimes, we fit the data in Fig.~\ref{fig:fig2}b to a simplified version of a theoretical  model. This model has been previously suggested for the unbinding rates of transcription factors (i.e., nucleic-acid binding proteins)  from single  binding sites along double stranded (ds)DNA~\cite{Kamar:2017dd,Sing:2014dz}. The model   assumes that unbinding of a multimeric ligand occurs in  a multi-step process. Each step can have its own salt dependence. At the first step, the ligand partially unbinds. At the later steps, the partially dissociated ligand desorbs gradually into bulk solution. For the sake of simplicity and to minimize the number of free fit parameters, here we assume  a two-step fashion and that only the first step has a salt dependence. This assumption is consistent with a previously reported analyses of the experimental data~\cite{Kamar:2017dd}. Thus, the off-rate can be expressed as the sum of two reaction times as
\begin{equation}
 k_{\mathrm{off}} =   \left( \frac{1}{\alpha c_\mathrm{s}^z} + \frac{1}{ k_{\mathrm{sat}} } \right)^{-1},
 \label{eq:fit_rates}
\end{equation}
where $\alpha, z,  k_{\mathrm{sat}}$ are free fit parameters to be determined for the implicit and explicit data sets separately. While the first term in parentheses in Eq.~\ref{eq:fit_rates} has a power-law dependence on the excess salt concentration, $c_\mathrm{s}$, with an exponent $z$, the second term in parenthesis   accounts for a salt-independent (saturation) process. 
Fitting our data to Eq.~\ref{eq:fit_rates}  gives $z_{\mathrm{exp}} = 1.35 \pm 0.02$  for the explicit case, which is slightly lower than that for the implicit case, $z_{\mathrm{imp}} = 1.46 \pm 0.04$. 

The finding that $z_{\mathrm{imp}}>z_{\mathrm{exp}}$  suggests 
a convergence  of the off-rates as $c_\mathrm{s} \rightarrow 0$ (or  as $\kappa^{-1} \rightarrow \infty$). However, we should note that even at vanishing  excess salt concentrations,   the counterions of the ligands and binding-sites are  present in the explicit simulations, unlike the implicit case. The exponent $z$ is often considered to correspond to the number of univalent salt ions replacing the ligand upon dissociation~\cite{Manning:1978fy,Record:1976vk, Anderson:1982wl}. In our case, the explicit ion simulations surprisingly lead to a similar exponent  with the  implicit cases, for which neither ions nor their excluded volumes are in present. This suggests that in our simulations the exponent $z$ may be rather a result of mixed effect of ionic entropy and electrostatic screening of the Coulomb interactions between the binding sites and the ligands. We will revisit this topic in the Conclusion section.

The saturation rates $ k_{\mathrm{sat}}$  obtained by fitting the off-rates to Eq.~\ref{eq:fit_rates}  
(Fig.~\ref{fig:fig2}b) 
differ roughly factor of  30 (i.e., $k_{\mathrm{sat}}^{\mathrm{imp}} =0.0058\; \tau^{-1}$ vs.  $k_{\mathrm{sat}}^{\mathrm{exp}} =0.0002\; \tau^{-1}$).  At high salt concentrations,  the electrostatic interactions between the charges  significantly weaken, thus, the difference in the saturation rates can be attributed to the contributions related to the degrees of freedom of the ions (e.g., ionic correlations and translational entropy), which are present only in the explicit-ion case.  Ion-ion correlations near the binding site are stronger~\cite{Sushko:2016fy} and thus they promote the bound ligand in favor of weaker correlations in bulk. In addition, since translational entropy of the ions decreases near the binding site, this entropy component also favors a bound ligand to maximize entropy~\cite{Misra:1994ii}.
%
%In addition, the highly concentrated ionic atmosphere near the binding site  can slow down the diffusion of the bound ligands compared to less dense bulk. We will discuss this issue more on the discussion section.
%

We note that while we see saturation behaviors  for  the off-rates in  both cases above  $c_\mathrm{s}>100\;$mM in Fig.~\ref{fig:fig2}b, the plateau regime in the explicit simulations appears at a slightly smaller threshold salt concentration (i.e., $c_\mathrm{s}^* = ( k_{\mathrm{sat}}/\alpha)^{1/z}$); the onset of the saturated regime in the implicit case  manifests itself at a roughly factor of $\sim 2$ more saltier solution.  This implies that the effects beyond the screening of electrostatic forces play a role in the salt-dependent behavior of unbinding in Fig.~\ref{fig:fig2}b.

To  further demonstrate the difference between the implicit and explicit cases,  in the inset of Fig.~\ref{fig:fig2}b, we show the ratios of the fitted functions of the off-rates. The inset demonstrates that the difference between the implicit and explicit cases is not simply a constant and has rather a salt dependent profile. At $c_\mathrm{s} \gtrsim 100\;$mM, the difference  increases gradually and  reaches a plateau. However, the difference between the off-rates decreases almost to a factor of 5 as the salt concentration decreases to $c_\mathrm{s} =5$ mM.  This is indeed expected since at vanishing salt concentration, neither entropic nor enthalpic effects of the ionic atmosphere are  present, and the  dissociation kinetics is determined by contact energy between the ligand and binding site. The electrostatic component of this energy increases with decreasing salt concentrations (i.e.,  the Debye length increases).

\subsubsection{On-rates only slightly differ for explicit and implicit salt}
\begin{figure}
\includegraphics[width=9cm,viewport= 0 0 740 530, clip]{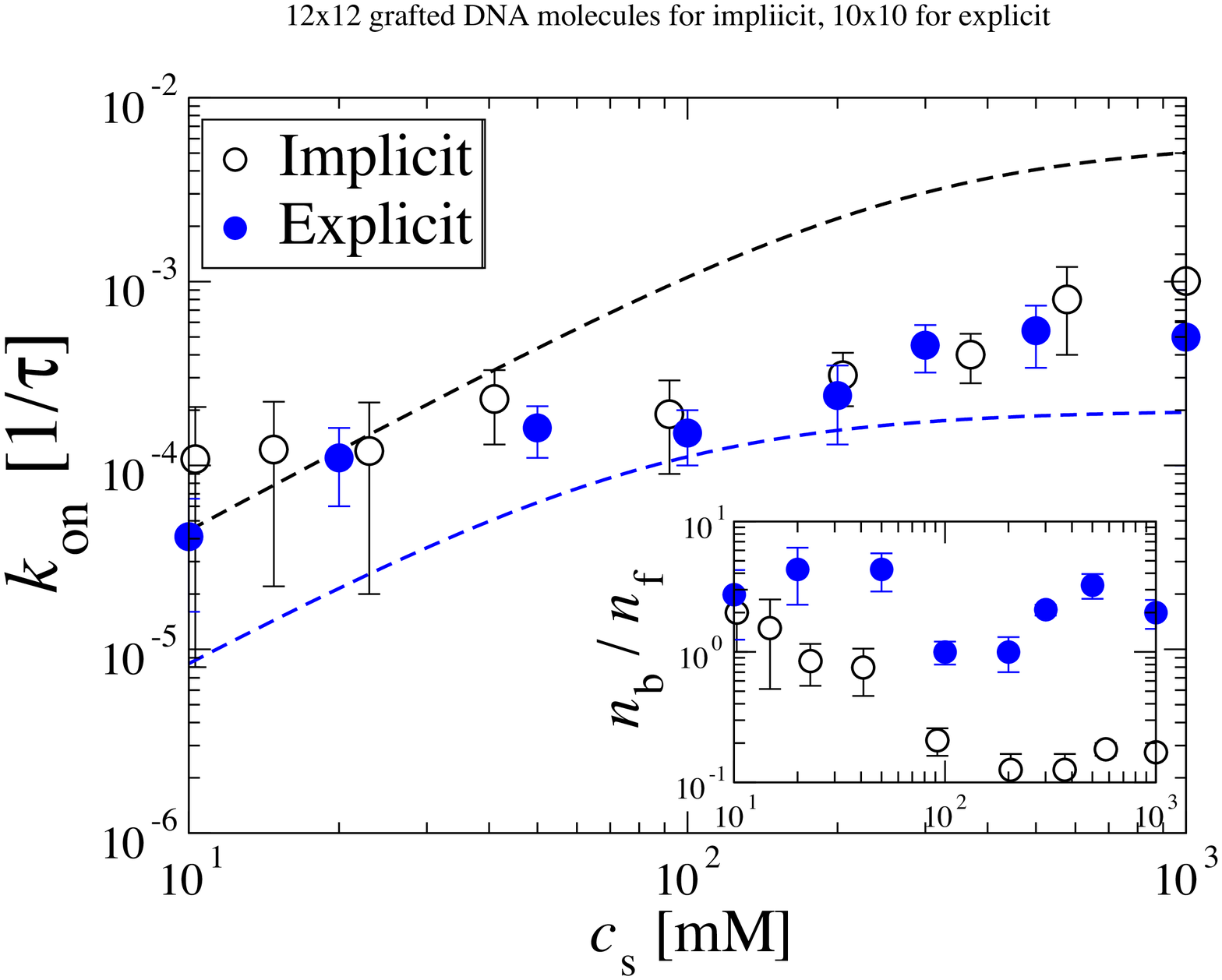}
\caption{The on-rates obtained from Eq.~\ref{eq:kon} for explicit and implicit electrostatics cases are shown as a function of excess salt concentration. The dashed curves show the off-rates of Fig.~\ref{fig:fig2}. The scale of y-axis is kept similar to that in Fig.~\ref{fig:fig2} for comparision. The inset shows the ratios of the number of bound and free ligands at respective steady states.
}
\label{fig:fig2b}
\end{figure}

Our simulation trajectories also allow us to compute the ligand on-rate $k_{\mathrm{on}}$, allowing us
check if the implicit electrostatics affects binding kinetics. 
We achieve this by considering the part of the trajectories, in which binding and unbinding events reach their quasi-steady states so that the on-rate can be expressed via a detailed balance expression as
\begin{equation}
 k_{\mathrm{on}} =   k_{\mathrm{off}}  \frac { n_{\mathrm{b}} } { n_{\mathrm{f}}},
 \label{eq:kon}
\end{equation}
where $n_{\mathrm{b}}$ and $n_{\mathrm{f}}$ are the number of  bound and free ligands at time $t > 1/k_{\mathrm{off}}$, respectively. In Fig.~\ref{fig:fig2b}, we show the calculated on-rates as a function of the excess salt concentration. For comparison, we also add the fit curves representing the off-rates in Fig.~\ref{fig:fig2}b for implicit and explicit cases. Unlike the off-rates, the on-rates do not show significant difference between the implicit and explicit treatments of the ions. According to Eq.~\ref{eq:kon}, this requires that on average there should be more ligands bound in the explicit case. As we show in the inset of Fig.~\ref{fig:fig2b}, the ratio $n_{\mathrm{b}}/n_{\mathrm{f}}$ differ between the implicit and explicit cases; there are more bound ligands in the explicit case, but the corresponding off-rates are smaller compared to the implicit case.
Therefore the change from explicit to implicit ions does not only change the unbinding kinetics, but also changes
the equilibrium binding site occupation probability, and therefore the binding free energy.

In Fig.~\ref{fig:fig2b}, we have lower error bars particularly at high salt concentrations, for which the sampling is better due to the higher frequencies of the unbinding events (i.e., shorter lifetimes). For the same reason,  the error bars in Fig.~\ref{fig:fig2b}  at low salt concentrations are larger due to the computational cost of  producing data when the electrostatic interactions are stronger. 

%
%
%\onecolumngrid
\begin{figure*}[t]
\centering
\includegraphics[width=16cm,viewport= 0 0 750 300, clip]{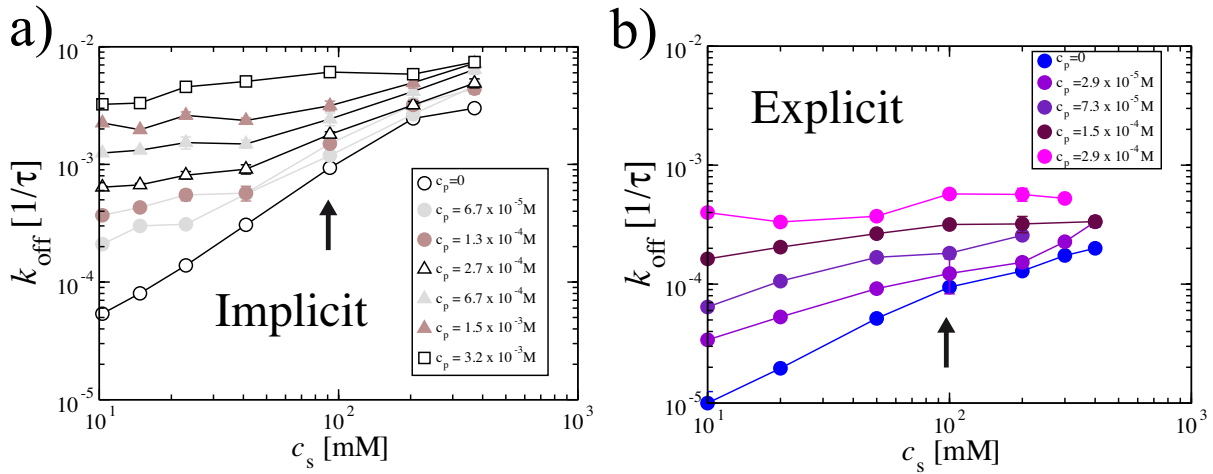}
\caption{The off-rates obtained from a) implicit  and b) explicit electrostatic simulations as a function of salt concentration  for various  free ligand concentrations. In the explicit simulations, all ions are modeled, whereas in the implicit case, a mean-field potential (Eq.~\ref{eq:Yukawa}) takes care of all ionic effects. The arrows indicate the salt concentrations focused in detail in Fig.~\ref{fig:fig4}. The data points are connected to guide the eye. }
\label{fig:fig3}
\end{figure*}
%\twocolumngrid
%
%
%%%%%%%%%%%%%%%%%%%%%%%%%%%%%%%%%%%%%%%%%%%%%%%
\subsection{Facilitated dissociation}
\subsubsection{Presence of binding ligand in solution weakens the effect of salt}

In experiments where protein unbinding from single binding sites was studied, a high  concentration of free ligands (proteins) weakened the salt-dependence of the off-rates~\cite{Graham:2011cy,Kamar:2017dd,Giuntoli:2015jf}. In our simulations, we also investigated the effects of the electrostatic scheme on FD by introducing a fixed concentration of unbound ligands in the solution (see Fig.~\ref{fig:fig1}).  In doing so, interplay between univalent ions and free cationic ligands (with valency +4) on the off-rates can be revealed. 
In Fig.~\ref{fig:fig3},  we show the off-rates of the ligands  as a function of salt concentration for various free ligand concentrations, $c_\mathrm{p}$, for the cases where we model electrostatics implicitly and explicitly. Fig.~\ref{fig:fig3} reveals that, regardless of the electrostatic treatment, if the number of free ligands  is increased, the off-rates become less sensitive to the changes in the salt concentration. This result is in accord with previous experiments~\cite{Kamar:2017dd}. As  $c_\mathrm{p}$ is increased, there are more ligands competing for the binding sites, hence,  the FD effects are more dominant. According to the data presented in  Fig.~\ref{fig:fig3}, this increasing competition weakens the salt-dependent steps of the unbinding process significantly~\cite{Kamar:2017dd}.

In  Fig.~\ref{fig:fig3}a and b, as  $c_\mathrm{p}$ is increased, the slopes in logarithmic  $k_{\mathrm{off}}$ versus $c_\mathrm{s}$ plots   decrease and eventually at high enough $c_\mathrm{p}$ values, respective plateau regimes appear. This behavior is qualitatively in agreement with the previous experiments~\cite{Kamar:2017dd}.
However, in the explicit case (Fig.~\ref{fig:fig3}b), the plateau regime appears at  smaller ligand concentrations with much slower off-rates compared to the implicit case (Fig.~\ref{fig:fig3}a);  almost one order of magnitude more ligand concentration is needed in the implicit simulations to reach the salt-independent off-rates (for instance, compare open squares in Fig.~\ref{fig:fig3}a and pink circles in Fig.~\ref{fig:fig3}b). These results suggest that  for the FD effect to take place,  an increase in steric  interactions on the binding site is essential~\cite{Tsai:2016gj, Kamar:2017dd}. In the explicit case, the presence of salt ions contribute this effect further by possibly increasing ion-binding site interactions~\cite{Misra:1994tg}, and thus, the plateau regime emerges at lower ligand concentrations compared to the implicit case. 

A quantitative comparison of the off-rates obtained from the implicit and explicit simulations in Fig.~\ref{fig:fig3} shows that at similar ligand concentrations, the off-rates for both cases are quite similar at low salt limits. For instance, at around $c_\mathrm{p} \approx 3 \times 10^{-4}\;$M, the off-rates are near $k_{\mathrm{off}} \approx 10^{-3}\;1/\tau$. This again suggests  that once ligand concentration is high enough, they  can replace the low concentration of salt ions near the binding sites and lead to similar kinetics rates for both cases.
 
\subsubsection{Facilitated dissociation is slower for explicit ions than for implicit ions}

\begin{figure}
\centering
\includegraphics[width=9cm,viewport= 0 0 750 550, clip]{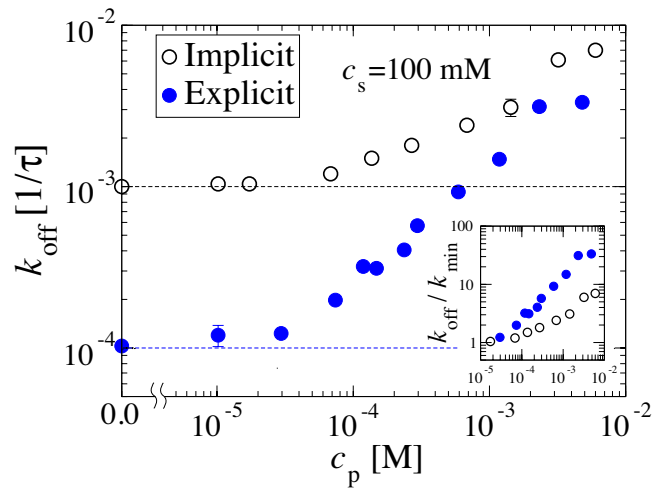}
\caption{The off-rates as a function of  free ligand concentration for implicit and explicit  cases are shown  at an excess salt concentration of $c_\mathrm{s} \approx 100\;$mM or screening length corresponding equivalent salt content. The vertical lines are to donate the  limit of $c_\mathrm{p} \rightarrow 0$. Inset shows the same data but rescaled by the minima of the off-rates at vanishing values of $c_\mathrm{p}$.}
\label{fig:fig4}
\end{figure}

The effect of electrostatic scheme on the FD can be seen more clearly in Fig.~\ref{fig:fig4}, where we show the off-rates for  implicit and explicit simulations as a function of ligand concentration at a  salt concentration of $c_\mathrm{s} \approx 100\;$mM. At low ligand concentrations, the presence of explicit ions considerably decreases the off-rates by almost one order of magnitude, as seen in Fig.~\ref{fig:fig4}. As the free ligand concentration, $c_\mathrm{p}$, is increased, deviations from the spontaneous dissociation rates (indicated by vertical dashed lines in Fig.~\ref{fig:fig4}) in both explicit and implicit cases become more drastic.  This can be seen more clearly in the inset of  Fig.~\ref{fig:fig4}, where we  show the same data sets but rescaled by the values of the spontaneous dissociation rates. The data in the inset demonstrates that in both cases, the deviations from the respective spontaneous-dissociation rates gradually increase. Also note that the deviations obey different slopes, thus, it is not possible to obtain a master curve for the two sets of  off-rates.

In  Fig.~\ref{fig:fig4}, as  $c_\mathrm{p}$ is further increased towards $c_\mathrm{p}  \approx 10^{-3}$ M, we observe  that the difference between implicit and explicit treatment of electrostatics enters a decreasing trend.  This behavior of the off-rates may be considered as result of a transition from  dilute to semidilute solution with increasing ligand concentration. In semidilute regime, excluded volumes of ligands can overlap even in bulk. We can calculate the critical concentrations, $c_\mathrm{p}^*$, for the ligands in various conditions. For our solution of ligand chains composed $p=4$ beads,   $c_\mathrm{p}^*$ on the scaling level  is $c_\mathrm{p}^* \approx  p /(p \sigma)^3 \approx 10^{-2}$ M with $\sigma =0.7$ for a strongly stretched conformation. In good-solvent conditions,  $c_\mathrm{p}^* \approx  p /(p^{3/5} \sigma)^3 \approx 10^{-1} $ M again with $\sigma =0.7$  nm. Both of these critical  concentrations are much larger  than the concentration, at which the off-rates for implicit and explicit cases meet in Fig.~\ref{fig:fig4}  (i.e., $c_\mathrm{p} \approx 10^{-3}$ M).  Thus, the convergence is not due to crowding of ligand chains in bulk phase.

One explanation for proximity of the off-rates in Fig.~\ref{fig:fig4} at around $c_\mathrm{p} \approx 10^{-3\;}$ M for the implicit and explicit cases  is that as the number of ligands competing for the same site increases, molecular overcrowding around the binding site by the ligands can  deplete the ions from the  binding site,  and thus, lead to similar off-rates regardless of the electrostatic treatment. If we calculate the radial distribution functions between the site and the ligand at $c_\mathrm{p} \approx 10^{-3\;}$ M (see SI Figure S2), indeed visually there is almost  no difference between the two cases; the binding site is equally crowded by the ligands in both implicit and explicit cases at high ligand concentrations.

Interestingly,   in  Fig.~\ref{fig:fig4}, at  $c_\mathrm{p} > 10^{-3}$ M, the difference between the off-rates obtained for the  implicit and explicit cases tend to  increase again. While the onset of the saturation  is  clearer in the explicit  simulations (this result will be supported by additional data  in the following section), for the implicit case, the saturation (if any) does not emerge clearly at the concentration range we study here. 
We  note that at very high ligand concentrations, the ligands themselves  can act like multivalent ions. Combined with explicit univalent ions, these ligands can yield  saturation values lower than those in the implicit case, similar to that observed in Fig.~\ref{fig:fig2}b.

%%Indeed, in the next section, we will see that  various non-electrostatic interaction strength will lead to a similar %%%effect at high ligand concentrations.
%
We underline that  as the concentration of a polyelectrolyte solution is increased to $c_\mathrm{p} \gg c_\mathrm{p}^*$ at high-salt conditions, the solution properties, such as effective viscosity, can change drastically~\cite{Muthukumar:1997tu}. Thus, at higher ligand and salt  concentrations that we did not consider here due to their low relevance to biological systems, the effects of implicit and explicit treatments on the off-rates can be more complex.

%
%
%
%%%%%%%%%%%%%%%%%%%%%%%%%%%%%%%%%%%%%%%%%%%%%%%%%%%%%%
\subsubsection{Short-ranged non-electrostatic interactions modulate the rate of facilitated dissociation}

In the previous subsections, we showed that use of explicit ions versus implicit ions
modifies the unbinding kinetics by considering various salt and ligand concentrations.
Those computations were done at a fixed value of non-electrostatic short-ranged interaction.
Biological ligands (proteins) can have specific interactions with binding sites, 
and that specificity can be varied appreciably, e.g., with DNA sequence. 
In this subsection we vary the strength of the non-electrostatic interactions and focus on the
cumulative effect of electrostatic and non-electrostatic interactions on unbinding kinetics during FD.

%electrostatic originated interaction between the DNA and binding proteins, interactions that are specific to the protein  contribute to the kinetic of unbinding events. The electrostatic and (non-electrostatic) specific interactions cumulatively construct an energy barrier or barriers that will control the frequency of unbinding events.
%
%In our implicit simulation, only electrostatic pair interactions (i.e., electrostatic enthalpy)  are modeled. In our explicit-ion simulations, the ionic atmosphere provides entropic components as well.

In our simulations, we control the non-electrostatic binding energy between the ligands and binding sites by varying the strength of the attractive potential in Eq.~\ref{eq:wca}. This attraction models enthalpic  interactions between the ligands and the binding sites (e.g., between proteins and nucleic acid chains). In general, increasing  the non-electrostatic attraction requires longer computational times to simulate complete titration kinetics because it shifts  off-rates downwards without changing the salt dependence (see SI Figure S3). However, high non-electrostatic binding strengths also enhance the separation of bound and semi-bound states of the ligands. 
This allows clearer observation of FD~\cite{Sing:2014dz,Dahlke:2017bn}. 

\begin{figure}
\includegraphics[width=9cm,viewport= 0 0 750 550, clip]{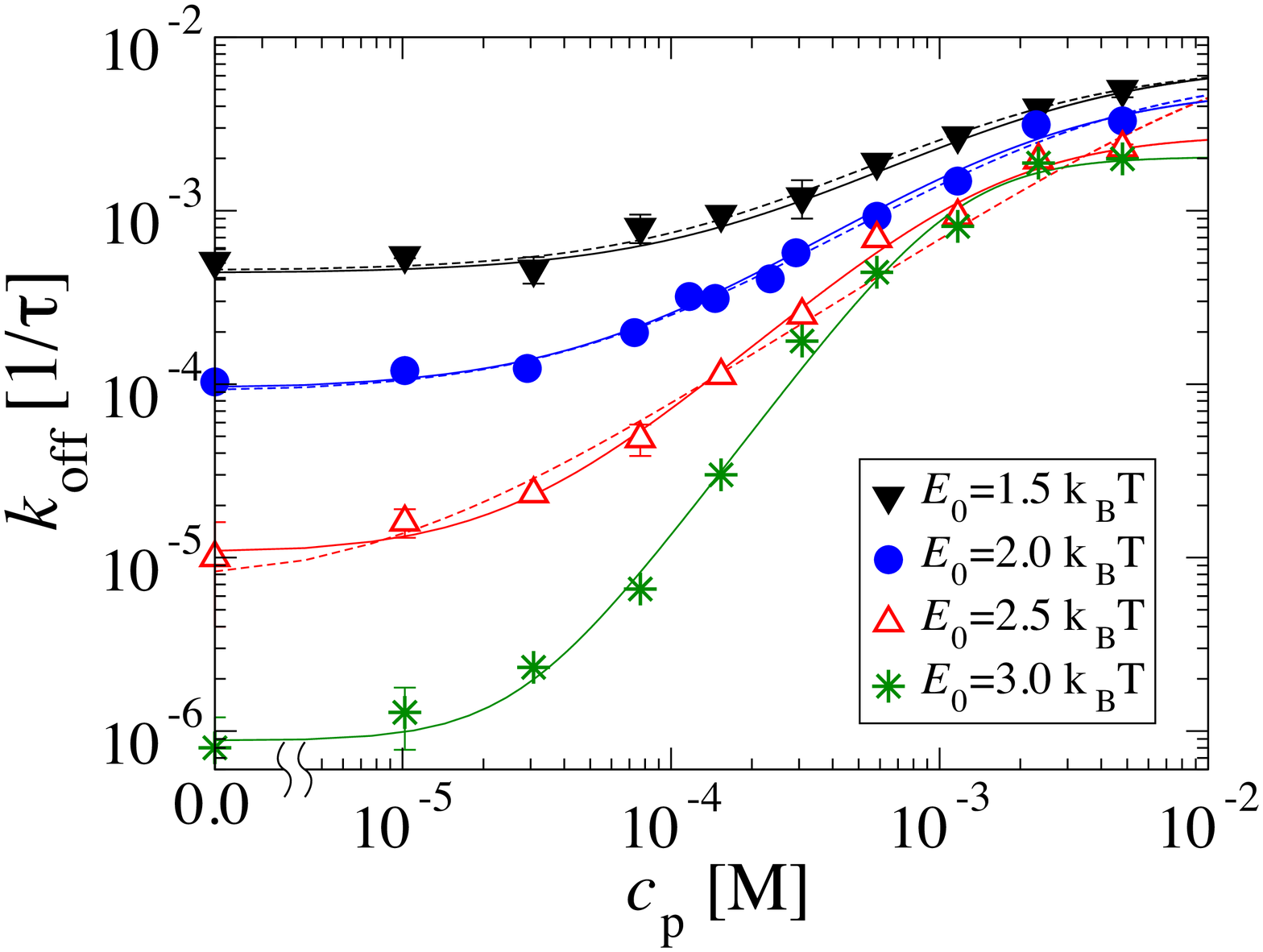}
\caption{The off-rates of the ligands as a function of free ligand concentration in solution are shown for various strengths of non-electrostatic binding energies for the explicit electrostatic cases at $c_\mathrm{s} = 100\;$mM. Dashed curves are fit functions obtained from Eq.~\ref{eq:fitscp} with $m=1$, whereas the solid curves are fit functions with $m$ being released as a fit parameter. 
} 
\label{fig:fig5}
\end{figure}

In Fig.~\ref{fig:fig5}, we show the simulation results for the explicit electrostatic simulations only for various non-electrostatic binding energies, since we have shown in the previous sections that implicit treatment cannot provide correct physical environment for unbinding. 
In the simulation, we vary the strength of the non-electrostatic binding energy per bead so that we can scan energy ranges around $~10\; k_{\mathrm{B}} T$ total, which is the typical molecular  binding energies~\cite{Erbas:2013ta,Tsai:2016gj,Sugimura:2006co}. Our simulations reveal various regimes for unbinding events.

As can be seen in  Fig.~\ref{fig:fig5}, at  low ligand concentrations, below $c_\mathrm{p}=10^{-5}\;$M, the off-rates are weakly dependent on  the concentration of free ligands in solution. In this  regime, the time needed for a bound ligand to spontaneously dissociate is relatively little compared to that for a free ligand to bind and destabilize the complex at the binding site. Interestingly, at  intermediate concentrations of excess ligands, there is a distinctive regime, whose slope  strongly depends on the non-electrostatic interaction strength between the ligand and the binding site (Fig.~\ref{fig:fig5}); as  the non-electrostatic binding strength is increased gradually from $E_0 = 1.5\;k_{\mathrm{B}} T$ to $E_0 = 3.0\; k_{\mathrm{B}} T$, the $c_\mathrm{p}$-dependence of the off-rate becomes steeper in the concentration range $c_\mathrm{p} \approx 10^{-5} -10^{-3}$ M.   As the concentration approaches  $c_\mathrm{p} \approx 10^{-3}$ M,  the off-rates for various strengths increase but tend to saturate at comparable values. A similar trend also emerges between the implicit and explicit simulations in Fig.~\ref{fig:fig4} (recall that all data in Fig.~\ref{fig:fig5} are obtained via explicit simulations).  Analogous universal behavior was proposed in a computational study of neutral ligands for a broader range of concentrations~\cite{Dahlke:2017bn}. We observe a convergence only at very high ligand concentrations in Fig.~\ref{fig:fig5} for various binding strengths.

To quantify the binding isotherms in Fig.~\ref{fig:fig5} obtained for various values of $E_0$,  we fit the data sets in Fig.~\ref{fig:fig5} to an equation in the form of 
\begin{equation}
k_{\mathrm{off}}  = \frac{c_\mathrm{p}^m + D }{A c_\mathrm{p}^m +B},
\label{eq:fitscp}
\end{equation}
where $A,B,D$ are the fit parameters to determine, and they correspond to low and high concentration limits in a few-state binding model (cf.~\cite{Kamar:2017dd}). The exponent $m$ usually  takes a value near unity for the 
familiar Langmuir binding isotherm in the absence of any cooperativity 
(i.e., binding events are not altered by already bound ligands)~\cite{Kamar:2017dd,Sing:2014dz}. The dashed curves in Fig.~\ref{fig:fig5} indeed are the fit functions obtained from Eq.~\ref{eq:fitscp} by settting $m=1$ (see SI Table 1  for the fit parameters). For the binding strengths  $E_0 \leq 2.5 \; k_{\mathrm{B}} T$, Eq.~\ref{eq:fitscp} can describe both low  and high concentration plateaus as well as the transition regimes successfully. However, for $E_0=3.0 \;k_{\mathrm{B}} T$, setting $m=1$ in Eq.~\ref{eq:fitscp} does not lead to a reasonable fit. If we release $m$ as a fit parameter in Eq.~\ref{eq:fitscp}, the data set for $E_0=3.0 \;k_{\mathrm{B}} T$  can also be fit to Eq.~\ref{eq:fitscp} but with a value of $m \approx 2$, whereas other data sets can still be described by $m \approx 1 $ (see SI Table 2  for the fit parameters).  

The fit value of $m>1$ for  $E_0=3.0 \;k_{\mathrm{B}} T$  in Fig.~\ref{fig:fig5}    suggests that high binding affinity can increase the binding of other ligands and increase cooperativity. With increasing binding affinity, the ligand bends the polymeric binding site stronger. Indeed, in the simulations, we observe this behavior qualitatively. Strong affinity towards the binding site, combined with local bending, can decrease rotational and translational mobility of the ligand on the binding site~\cite{Tan:2016ff}. Thus, a newly attached ligand has a higher chance to bind to the  site with minimum interference from the bound ligand.

 Eq.~\ref{eq:fitscp} also provides valuable information for the low and high concentration limits, $D/B$ and $1/A$, respectively.  While the high concentration saturation rate, $1/A$, decreases with increasing $E_0$, the variation among different strengths is not more than factor of 3 in  Fig.~\ref{fig:fig5}. Consistently, the difference in $1/A$'s of $E_0=3.0\;k_{\mathrm{B}} T$   and $E_0=2.5\; k_{\mathrm{B}} T$ is rather insignificant (see SI Table 1 and 2).  For the low concentration limits, the values of $D/B$  range between $\approx 10^{-6}$ and $10^{-1}\;1/\tau$ (see SI Tables 1 and 2), suggesting  exponentially decreasing off-rates with increasing $E_0$  at $c_\mathrm{s} = 100$ mM (i.e., $D/B \sim \exp (2 E_0 / k_{\mathrm{B}} T)$)~\cite{Sing:2014dz}.

Interestingly, the effect of non-electrostatic interactions at the intermediate ligand concentrations is also evident  in the implicit electrostatic simulations with the systematic  overestimation of the off-rates (see SI Figure S4). 
The similarities in the off-rates in the explicit and implicit as a function of solvated ligands confirm that  steric interactions between two or more ligands on the binding site are essential for FD process.

\section{Conclusions}

In this study, by using a generic polyelectrolyte model, we examined the effects of  implicit (mean field) 
versus explicit treatments of ionic components on  the dissociation kinetics of single ligands from polymeric binding sites. 
We studied the role of electrostatics in both spontaneous and facilitated disassociation mechanisms  at various salt and excess ligand concentrations. Our simulations revealed that a mean-field level treatment of electrostatics overestimates the unbinding rates compared to explicit treatment of ionic species (e.g., by considering their net charge and steric interactions). The explicit  ions  provide  translational entropy and ionic correlation contributions, both of which favor slower unbinding rates.  Furthermore, the trend in unbinding rates versus the ligand concentration data does not allow us to obtain a master curve, in which the implicit and explicit simulation data could be overlapped. This suggest that the explicit ions change the underlying physics of the FD.  We now discuss our major results in slightly more detail.

\subsection{Spontaneous dissociation kinetics are significantly different in explicit versus implicit ion cases}
Ionic effects are expected to play major roles in the unbinding kinetics 
of molecular ligands.  First of all, with increasing salt concentration, the Debye length decreases. The ionic correlations in bulk are more correlated at length scales smaller than the Debye length. At scales larger than the Debye length, Coulomb interactions are screened. Secondly,  entropic and enthalpic contributions of ionic atmosphere near the binding site and ligand can favor or inhibit dissociation, depending on the salt concentration. In our implicit simulations, only the screening effect of salt is available via manipulation of the Debye length in the pair potential given in Eq.~\ref{eq:Yukawa}. In our explicit simulations, where all the univalent  ions are considered explicitly,  charge-related and steric (excluded volume) effects of the ions are present. Hence, in addition to the screening effect, in the explicit case, the ions arrange themselves near the binding site and the ligand relative to the bulk  by balancing corresponding entropy and enthalpic energy components. Below we discuss these components and their role in the unbinding processes along with our simulation results.

\subsubsection{Salt dependence of unbinding  is determined by electrostatic screening  at low salt}
In the implicit simulation, the  binding affinity has two contributions;  a non-electrostatic attraction characterized by $E_0$, 
and an electrostatic Coulomb attraction between the oppositely charged binding site and the ligand. These two contributions favor a bound ligand. The non-electrostatic binding energy has no salt dependence (we assume that the salt does not change chain conformations drastically).  Contrarily, the electrostatic attraction  becomes weaker with increasing salt concentration.  As the electrostatic part of the  binding affinity decreases with increasing salt concentration, thermal fluctuations promote (spontaneous) dissociation of the ligand at $c_{\mathrm{s}} \lesssim 100\;$M. Thus, the slope of $\approx 1.4$ that we observe in the off-rate versus salt concentration plots below physiological salt concentrations in Fig.~\ref{fig:fig2}b can only be  due to the weakening of the electrostatic pairwise interactions (Eq.~\ref{eq:Yukawa}).  At $c_{\mathrm{s}}> 100\;$M, increasing salt concentration weakly change the electrostatic part of the attraction since $\kappa^{-1} < 1 \sigma$. As a result, the non-electrostatic binding strength, $E_0$,  controls the unbinding. It is important to note that ion concentration per Debye volume scales as $\kappa^3 \sim   c_\mathrm{s}^{3/2}$, which is somewhat  close to the exponents observed in Fig.~\ref{fig:fig2}b.
 
 At this point, we remind reader that on our polymeric binding sites,  the 1D charge density is set to $1/  \ell_{\mathrm{B}}$ (i.e., condensed ions are not considered explicitly)  for a  proper comparison between explicit and implicit electrostatic treatments since the screening effect of ``condensed'' ions is not well defined~\cite{delaCruz:1995vz}. 
 %Hence, the ionic effects we discuss here shall be considered for diffuse ion cloud. 
 %
We anticipate that adding ``condensed'' ions in the simulations would change the slope of the off-rates. In that case, the main contribution to the ion-release effect would be mainly due to ``condensed'' ions rather than relatively rapidly diffusing ions near the binding site, which we already include in our explicit-ion simulations. This  suggests that in studying the kinetics of ligand unbinding, the contributions from condensed and diffuse ions should be considered separately to comprehend ionic effects in more detail.  
 
\subsubsection{Quantitatively lower unbinding rates in explicit simulations underline transnational entropy and correlations effects of ions} 

In the explicit simulations, the non-electrostatic binding affinity and electrostatic the Coulomb attraction also promote a bound ligand.  The quantitatively lower off-rates obtained in the explicit simulations relative to those in the implicit cases (Figs.~\ref{fig:fig2} and~\ref{fig:fig3}) suggest that additional effects due to the explicit ions  extends the lifetime of the bound ligand on the binding site.

On a naked binding site, an ion cloud interact with the binding site by forming transient ion pairs. Once a ligand binds to the site, the ionic cloud is distorted by the ligand. If the bound  ligand  compensates the loss ion pairs energetically, the corresponding energy change between bound and unbound states can be ignored~\cite{Record:1976vk}. If this is not the case (i.e., the loss ions pairs are not compensated upon binding), the unbound state can be favored by the interactions between the surrounding ion cloud  and the binding site~\cite{Misra:1994ii}. The strength of this effect intensifies with increasing salt concentration since more ions can interact with the binding site.  Note that  in the implicit simulations, there are no ions to compensate electrostatic energy loss once a ligand desorbs  at low excess ligand concentrations. 
%Hence, the electrostatic attraction between structural charges promotes binding.

In the explicit simulations, the ionic correlations  near the binding site (and ligand) are ``tighter'' than in the bulk  as a due to the electrostatic potential induced by the negatively charged binding site.   The free energy associated with these ionic correlations favors a bound ligand~\cite{Misra:1994tg}, since dissociation of the bound ligand relocates more number of  ions  from bulk to nearby the binding site. This effect grows with increasing salt concentration since more ions must arrange themselves near the binding site once the ligand unbinds~\cite{Misra:1994ii},

Another effect that can slow down the off rates in the explicit simulations  is the translational entropy of the ions. Ideally  ions tend to diffuse away from the structural charges to increase their translational entropy. A bound ligand disperses a number of ions near the structural charges to bulk. Contrarily, upon unbinding, the ions need to replace the unbound ligand  by sacrificing their translational  entropy. Thus, the  translational entropy gain of the ions in bulk works in the advantage of the bound ligand. This effect weakens as the salt concentration becomes more uniform throughout the solution~\cite{Misra:1994ii}.

Overall, the Coulomb attraction between the binding site and the ligand is present for both explicit and implicit cases and favors a bound ligand. In the explicit-ion simulations, both translational entropy and correlation contributions also favor a bound ligand, and they are absent in the implicit simulations. These two ionic contributions in the explicit case extend the lifetime of the ligands on the binding site, and thus, lead to much lower off rates. Interestingly, with increasing salt, the translational entropy contribution decreases whereas the contributions due to the correlations increases~\cite{Misra:1994ii}. Moreover, in the explicit simulations, the interactions between the ion cloud and the structural charges can also favor or inhibit a bound state depending on the balance of the electrostatic energy upon unbinding. It is possible that  these ion-related contributions cumulatively lead to the behavior observed for the ratios of the off rates  in  the inset of in Fig.~\ref{fig:fig2}b obtained for the implicit and explicit cases.

\subsection{Interplay of ions and free ligands in facilitated dissociation } 

In the FD simulations, at low ligand concentrations the mechanism mentioned above for the spontaneous dissociation still controls the unbinding kinetics. At intermediate excess ligand  concentrations,  ideally a bound ligand is replaced by other ligand  that has been previously in bulk. As a result, the energetic contributions we discuss above should not change as long as the ligands are identical.  In this case, the ligand dissociation is determined by the repulsive and steric interactions  between two (or more) ligands on the same binding site. As the non-electrostatic binding affinity, $E_0$, is increased, it is harder for the bound ligands to kick one another out of the binding site (Fig.~\ref{fig:fig5}).
 
 For both implicit and explicit cases, we obtain salt-independent regimes as the excess ligand concentrations is increased (Fig.~\ref{fig:fig3}). This result is consistent with the previous experiments, where the salt-dependence of the off-rate of the DNA binding proteins becomes weaker with increasing protein concentration in solution~\cite{Kamar:2017dd}. However, for the implicit case, one order of magnitude more ligand concentration is required to observe this regime (the off-rates in the implicit case are still faster compared to those in the explicit case).  
This suggests that the salt-independent regime requires a highly competitive binding site environment provided either by the ions or excess ligands competing to replace the bound ligand.   In the explicit simulations,   excluded volumes of the ions near the binding site enhance this effect by possibly promoting more partially bound ligands. These partially bound ligands are more prone to FD. For the implicit simulations,  however,  the excluded volumes  of the ions  are not present, and thus, a higher number of ligands per binding site is needed to observe the FD effect. 

 At high ligand concentrations, in Fig.~\ref{fig:fig5} the  off-rates exhibit a tendency towards a common saturation plateau, regardless of the strength of the non-electrostatic binding, $E_0$.  Such an effect requires that any contribution due to the non-electrostatic binding energy separating the bound and unbound states  weakens or disappears.  Such a scenario can arise if the local volume fraction of the monomers near the binding site  reaches a value of unity (i.e., the limit of polymer melts). Under this condition,the  two-body interactions between the monomers can be  screened out (i.e., second virial coefficient approaches  zero)~\cite{Rubinstein:2003vd}. In other words, the solvent quality  changes at the local level and renders the non-electrostatic binding strength irrelevant. Indeed, a comparison of the radial distribution functions between the binding sites and the ligands shows that the local ligand concentrations on the binding site  for $E_0>1.5 \;k_\mathrm{B} T$ approach each other  at high enough values of $c_\mathrm{p}$  (see SI Figure S5).

\subsection{Explicit nature of ionic atmosphere is essential for modeling unbinding kinetics}

In the salt concentration range we consider (i.e., $c_\mathrm{s} = 5 -1000\;$mM), the distribution of explicitly modeled ions around polymeric binding site can lead to non-uniform screening effects. The small ions can destabilize the complex non-trivially and lead to local electrostatic screening. Furthermore, the bound cationic ligands alters the electrostatic potential of the binding site by  bringing co-ions close to the binding site. Such effects cannot be described in the implicit electrostatic simulations, since the mean-field nature of Eq.~\ref{eq:Yukawa} provides an electrostatic screening regardless of the conformation and relative position of the components (i.e., the Debye length is a function of total salt concentration).

In the on-rates we calculate via detailed balance (Fig.~\ref{fig:fig2b}), we observe that the difference between the implicit and explicit cases is rather small.  This suggest that the large difference in the off-rates reported here is not a simple consequence of procedure of coarse-graining ionic components.   The inset in Fig.~\ref{fig:fig2}, where we show the ratio of the off-rates obtained via implicit and explicit simulations, also supports this argument, since this ratio is neither a simple rescaling constant nor  a linear function of the salt concentration.

\subsection{Implicit simulations can underestimate contact energies}
In the implicit simulations, the contact energy between two beads (forming either the ligands or the binding site) at contact can be weaker due to mean-field nature of Eq.~\ref{eq:Yukawa}. This is because, by definition, Eq.~\ref{eq:Yukawa} gives a rescaled contact (electrostatic) energy of $\ell_{\mathrm{B}} / \sigma \exp (-\kappa \sigma) <\ell_{\mathrm{B}} / \sigma $ (recall that $\ell_{\mathrm{B}} / \sigma$ is the  electrostatic contact energy rescaled by $k_{\mathrm{B}} T$ between two beads of identical sizes of 1 $\sigma$ at zero salt).  The lower contact energy can also accelerate unbinding

To test how the electrostatic contact energy changes the off-rates, we perform additional test simulations  by increasing the contact energy in the  implicit simulations. We achieve this by shifting the decay term in Yukawa potential (i.e., $\ell_{\mathrm{B}} / r \exp (-\kappa ( r - \sigma)  = \ell_{\mathrm{B}} / r $) so that at $r=\sigma$, the implicit and explicit simulations give similar contact energies (here we ignore the effects of the  ions surrounding the ligand and binding site). In those simulations, we obtain off-rates that are  lower than those in the explicit-case (see SI Figure S6). While this test simulations can demonstrate the effect of the contact energy in unbinding, an arbitrary rescaling of  the mean-field potential between the ligand and binding site is necessary to exactly match the results of the explicit-ion simulations.

%This suggests that inevitably all mean-field type  electrostatic approaches  fail to describe the kinetics between the ligand and host correctly since   Note that,  in explicit-ion simulations, individual salt ions alter the Coulombic pairwise interactions by physically placing themselves between and near  the structural charges  by overcoming the steric interaction.

%The electrostatic and (non-electrostatic) specific interactions cumulatively construct an energy barrier or barriers that will control the frequency of unbinding events. The rate for unbinding events can be expressed as a $k_{\mbox {\tiny off}}  \propto \exp (-\Delta G / k_{\mathrm{B}} T)$, where $\Delta G$ is the dimensionless energy  barrier separating bound and unbound states.

\subsection{Non-uniform dielectric constant can accelerate the binding}
In our coarse-grained MD simulations, we use a continuum solvent model with a constant background dielectric constant. In a more realistic, orientation of water molecules near the binding sites, polar groups composing the ligand itself, and the salt concentration would influence the local dielectric constant and thus, ionic distributions~\cite{Honig:1995tt,Bonthuis:2011cg}. Indeed, recently it has been shown that ionic mobility in salt solutions can be described more accurately with a non-uniform dielectric constant~\cite{Fahrenberger:2015ku}. Similar effects can alter unbinding rates by pushing the ions from the binding site or increasing diffusion coefficient within the ionic atmosphere. Nonetheless, these and similar effects can enhance the importance of explicit treatment of ionic atmosphere in studies, where kinetics of biomolecules are considered.

\subsection{Distribution  of binding sites and chromatin in vivo} 
In the setup we consider, the binding sites arranged on a 2D surface. A 3D distribution of binding sites is  expected  in the case of protein specific binding sites along chromatin confined by nuclear envelope or bacterial cell membrane. Since, as we show here, the off-rates depend on the local  interactions and ionic distributions around binding sites, we do not expect that the distribution of binding sites will change the results we present here as long as the distance between the binding sites is larger the Debye length. However, in a highly concentrated  polyelectrolyte matrix,  electrostatic potentials of constituting chains can broaden the ion distributions around neighboring chains~\cite{Li:2016cu}. In such cases, exchange rates can be a function of  local polymer conformation~\cite{Parsaeian:2013iu}.

\subsection{Quantitative modeling of protein-dsDNA interactions} 
Our calculations have been motivated by consideration of experiments with proteins binding to dsDNA binding sites.
One might ask how accurately our MD model can be used to physically describe the very high charge density of DNA.
In our simulations, we set the size of each bead composing the binding sites and ligands to 0.7 nm. By setting the size of the beads to $\approx 1.4$ nm and their charges to $-2$ without changing the size of the ions, a model mimicking binding sites along dsDNA  is obtained. The simulations with those parameters do not lead to any drastic difference. At low salt concentrations, the off-rates have a weaker dependence on salt (see SI Figure S7). In general, we should note that models like ours can be used to understand the fundamental physical principles of electrostatic effects. However, atomistic or less coarse-grained computational models are more suitable to study molecular aspects of DNA binding proteins~\cite{KnottsIV:2007gn,Uusitalo:2015eh}. We plan to use one of these models to further study unbinding kinetics in the near future.

%\textit{Multivalent ions:}

%Ionic interactions play a fundamental role in ligand dissociation from bindings sites  either by screening  electrostatic potentials between constituting species or by creating a competition between various entropic and entalphic components. 
%However, complexity of the problem does not allow a clear separation of these effects in most realistic scenarios. 
%

\subsection{Final remarks}

Our results indicate that condensed ions can have stronger effects in determining the slope of dissociation constant in the low-salt limit since we obtain qualitatively similar behaviors for our unbinding rates in the implicit and explicit simulations, in which only the effects of uncondensed charges are introduced a priori. The contributions from condensed and diffuse ions should be considered separately to infer ionic effects in unbinding kinetics.
Similarly,
in our facilitated dissociation simulations, the implicit treatment led to higher unbinding rates for various excess ligand concentrations at all salt concentrations we considered. The salt concentration has a much weaker effect on the facilitated dissociation due to depletion of ions from  binding sites.  At high ligand concentrations, at which binding site is overcrowded by ligands, we observe a universal saturation regime of facilitated dissociation, regardless of non-electrostatic binding strength or electrostatic treatment. This leaves the possible role  of nonidentical ligands in unbinding process as an open question for further investigations. 
Finally, our results also suggest cooperative binding of ligands for high non-electrostatic binding strengths, suggesting
a general mechanism whereby
any protein bound to a binding site can promote binding of a second protein near the same site, even without
direct interactions of the adjacent proteins.

%As a final remark, our model here with full electrostatics can be used to study more specific systems, such as protein dissociations from long or short nucleic acid chains, after rescaling related length and energy  scales accordingly and carefully.

\begin{acknowledgements}
%{\centering \textbf{Acknowledgements}}
AE acknowledges Edward J. Banigan and Martin Girard for their careful readings of the manuscript. The computational resources at Northwestern University (Quest) are greatly acknowledged.
This work was supported by the National Science Foundation through grants DMR-1611076  and DMR-1206868
and by the National Institutes of Health through grants U54-CA193419, U54-DK107980 and R01-GM105847.

% put your grant number here!!
\end{acknowledgements}

%\bibliographystyle{unsrt}
\bibliography{Erbas_etal}

\begin{thebibliography}{56}%
\makeatletter
\providecommand \@ifxundefined [1]{%
 \@ifx{#1\undefined}
}%
\providecommand \@ifnum [1]{%
 \ifnum #1\expandafter \@firstoftwo
 \else \expandafter \@secondoftwo
 \fi
}%
\providecommand \@ifx [1]{%
 \ifx #1\expandafter \@firstoftwo
 \else \expandafter \@secondoftwo
 \fi
}%
\providecommand \natexlab [1]{#1}%
\providecommand \enquote  [1]{``#1''}%
\providecommand \bibnamefont  [1]{#1}%
\providecommand \bibfnamefont [1]{#1}%
\providecommand \citenamefont [1]{#1}%
\providecommand \href@noop [0]{\@secondoftwo}%
\providecommand \href [0]{\begingroup \@sanitize@url \@href}%
\providecommand \@href[1]{\@@startlink{#1}\@@href}%
\providecommand \@@href[1]{\endgroup#1\@@endlink}%
\providecommand \@sanitize@url [0]{\catcode `\\12\catcode `\$12\catcode
  `\&12\catcode `\#12\catcode `\^12\catcode `\_12\catcode `\%12\relax}%
\providecommand \@@startlink[1]{}%
\providecommand \@@endlink[0]{}%
\providecommand \url  [0]{\begingroup\@sanitize@url \@url }%
\providecommand \@url [1]{\endgroup\@href {#1}{\urlprefix }}%
\providecommand \urlprefix  [0]{URL }%
\providecommand \Eprint [0]{\href }%
\providecommand \doibase [0]{http://dx.doi.org/}%
\providecommand \selectlanguage [0]{\@gobble}%
\providecommand \bibinfo  [0]{\@secondoftwo}%
\providecommand \bibfield  [0]{\@secondoftwo}%
\providecommand \translation [1]{[#1]}%
\providecommand \BibitemOpen [0]{}%
\providecommand \bibitemStop [0]{}%
\providecommand \bibitemNoStop [0]{.\EOS\space}%
\providecommand \EOS [0]{\spacefactor3000\relax}%
\providecommand \BibitemShut  [1]{\csname bibitem#1\endcsname}%
\let\auto@bib@innerbib\@empty
%</preamble>
\bibitem [{\citenamefont {Levy}\ \emph {et~al.}(2007)\citenamefont {Levy},
  \citenamefont {Onuchic},\ and\ \citenamefont {Wolynes}}]{Levy:2007ki}%
  \BibitemOpen
  \bibfield  {author} {\bibinfo {author} {\bibfnamefont {Y.}~\bibnamefont
  {Levy}}, \bibinfo {author} {\bibfnamefont {J.~N.}\ \bibnamefont {Onuchic}}, \
  and\ \bibinfo {author} {\bibfnamefont {P.~G.}\ \bibnamefont {Wolynes}},\
  }\href@noop {} {\bibfield  {journal} {\bibinfo  {journal} {Journal of the
  American Chemical Society}\ }\textbf {\bibinfo {volume} {129}},\ \bibinfo
  {pages} {738} (\bibinfo {year} {2007})}\BibitemShut {NoStop}%
\bibitem [{\citenamefont {Jones}\ \emph {et~al.}(2003)\citenamefont {Jones},
  \citenamefont {Shanahan}, \citenamefont {Berman},\ and\ \citenamefont
  {Thornton}}]{Jones:2003be}%
  \BibitemOpen
  \bibfield  {author} {\bibinfo {author} {\bibfnamefont {S.}~\bibnamefont
  {Jones}}, \bibinfo {author} {\bibfnamefont {H.~P.}\ \bibnamefont {Shanahan}},
  \bibinfo {author} {\bibfnamefont {H.~M.}\ \bibnamefont {Berman}}, \ and\
  \bibinfo {author} {\bibfnamefont {J.~M.}\ \bibnamefont {Thornton}},\
  }\href@noop {} {\bibfield  {journal} {\bibinfo  {journal} {Nucleic Acids
  Research}\ }\textbf {\bibinfo {volume} {31}},\ \bibinfo {pages} {7189}
  (\bibinfo {year} {2003})}\BibitemShut {NoStop}%
\bibitem [{\citenamefont {Record}\ \emph {et~al.}(1976)\citenamefont {Record},
  \citenamefont {Lohman},\ and\ \citenamefont {De~Haseth}}]{Record:1976vk}%
  \BibitemOpen
  \bibfield  {author} {\bibinfo {author} {\bibfnamefont {M.~T.}\ \bibnamefont
  {Record}}, \bibinfo {author} {\bibfnamefont {T.~M.}\ \bibnamefont {Lohman}},
  \ and\ \bibinfo {author} {\bibfnamefont {P.}~\bibnamefont {De~Haseth}},\
  }\href@noop {} {\bibfield  {journal} {\bibinfo  {journal} {Journal of
  Molecular Biology}\ }\textbf {\bibinfo {volume} {107}},\ \bibinfo {pages}
  {145} (\bibinfo {year} {1976})}\BibitemShut {NoStop}%
\bibitem [{\citenamefont {Anderson}\ and\ \citenamefont
  {Record}(1995)}]{Anderson:1995tg}%
  \BibitemOpen
  \bibfield  {author} {\bibinfo {author} {\bibfnamefont {C.~F.}\ \bibnamefont
  {Anderson}}\ and\ \bibinfo {author} {\bibfnamefont {M.~T.}\ \bibnamefont
  {Record}, \bibfnamefont {Jr}},\ }\href@noop {} {\bibfield  {journal}
  {\bibinfo  {journal} {Annual Review of Physical Chemisty}\ }\textbf {\bibinfo
  {volume} {46}},\ \bibinfo {pages} {657} (\bibinfo {year} {1995})}\BibitemShut
  {NoStop}%
\bibitem [{\citenamefont {Anderson}\ and\ \citenamefont
  {Record}(1982)}]{Anderson:1982wl}%
  \BibitemOpen
  \bibfield  {author} {\bibinfo {author} {\bibfnamefont {C.~F.}\ \bibnamefont
  {Anderson}}\ and\ \bibinfo {author} {\bibfnamefont {M.~T.}\ \bibnamefont
  {Record}, \bibfnamefont {Jr}},\ }\href@noop {} {\bibfield  {journal}
  {\bibinfo  {journal} {Annual Review of Physical Chemisty}\ }\textbf {\bibinfo
  {volume} {33}},\ \bibinfo {pages} {191} (\bibinfo {year} {1982})}\BibitemShut
  {NoStop}%
\bibitem [{\citenamefont {Rouzina}\ and\ \citenamefont
  {Bloomfield}(1998)}]{Rouzina:1998bw}%
  \BibitemOpen
  \bibfield  {author} {\bibinfo {author} {\bibfnamefont {I.}~\bibnamefont
  {Rouzina}}\ and\ \bibinfo {author} {\bibfnamefont {V.~A.}\ \bibnamefont
  {Bloomfield}},\ }\href@noop {} {\bibfield  {journal} {\bibinfo  {journal}
  {Biophysical Journal}\ }\textbf {\bibinfo {volume} {74}},\ \bibinfo {pages}
  {3152} (\bibinfo {year} {1998})}\BibitemShut {NoStop}%
\bibitem [{\citenamefont {Manning}(1978)}]{Manning:1978fy}%
  \BibitemOpen
  \bibfield  {author} {\bibinfo {author} {\bibfnamefont {G.~S.}\ \bibnamefont
  {Manning}},\ }\href@noop {} {\bibfield  {journal} {\bibinfo  {journal}
  {Quarterly Reviews of Biophysics II}\ }\textbf {\bibinfo {volume} {2}},\
  \bibinfo {pages} {179} (\bibinfo {year} {1978})}\BibitemShut {NoStop}%
\bibitem [{\citenamefont {Privalov}\ \emph {et~al.}(2011)\citenamefont
  {Privalov}, \citenamefont {Dragan},\ and\ \citenamefont
  {Crane-Robinson}}]{Privalov:2011kw}%
  \BibitemOpen
  \bibfield  {author} {\bibinfo {author} {\bibfnamefont {P.~L.}\ \bibnamefont
  {Privalov}}, \bibinfo {author} {\bibfnamefont {A.~I.}\ \bibnamefont
  {Dragan}}, \ and\ \bibinfo {author} {\bibfnamefont {C.}~\bibnamefont
  {Crane-Robinson}},\ }\href@noop {} {\bibfield  {journal} {\bibinfo  {journal}
  {Nucleic Acids Research}\ }\textbf {\bibinfo {volume} {39}},\ \bibinfo
  {pages} {2483} (\bibinfo {year} {2011})}\BibitemShut {NoStop}%
\bibitem [{\citenamefont {Vander~Meulen}\ \emph {et~al.}(2008)\citenamefont
  {Vander~Meulen}, \citenamefont {Saecker},\ and\ \citenamefont
  {Record}}]{VanderMeulen:2008kq}%
  \BibitemOpen
  \bibfield  {author} {\bibinfo {author} {\bibfnamefont {K.~A.}\ \bibnamefont
  {Vander~Meulen}}, \bibinfo {author} {\bibfnamefont {R.~M.}\ \bibnamefont
  {Saecker}}, \ and\ \bibinfo {author} {\bibfnamefont {M.~T.}\ \bibnamefont
  {Record}, \bibfnamefont {Jr}},\ }\href@noop {} {\bibfield  {journal}
  {\bibinfo  {journal} {Journal of Molecular Biology}\ }\textbf {\bibinfo
  {volume} {377}},\ \bibinfo {pages} {9} (\bibinfo {year} {2008})}\BibitemShut
  {NoStop}%
\bibitem [{\citenamefont {Sugimura}\ and\ \citenamefont
  {Crothers}(2006)}]{Sugimura:2006co}%
  \BibitemOpen
  \bibfield  {author} {\bibinfo {author} {\bibfnamefont {S.}~\bibnamefont
  {Sugimura}}\ and\ \bibinfo {author} {\bibfnamefont {D.~M.}\ \bibnamefont
  {Crothers}},\ }\href@noop {} {\bibfield  {journal} {\bibinfo  {journal}
  {Proceedings of the National Academy of Sciences, U.S.A}\ }\textbf {\bibinfo
  {volume} {103}},\ \bibinfo {pages} {18510} (\bibinfo {year}
  {2006})}\BibitemShut {NoStop}%
\bibitem [{\citenamefont {Koblan}\ and\ \citenamefont
  {Ackers}(2002)}]{Koblan:2002hl}%
  \BibitemOpen
  \bibfield  {author} {\bibinfo {author} {\bibfnamefont {K.~S.}\ \bibnamefont
  {Koblan}}\ and\ \bibinfo {author} {\bibfnamefont {G.~K.}\ \bibnamefont
  {Ackers}},\ }\href@noop {} {\bibfield  {journal} {\bibinfo  {journal}
  {Biochemistry}\ }\textbf {\bibinfo {volume} {30}},\ \bibinfo {pages} {7822}
  (\bibinfo {year} {2002})}\BibitemShut {NoStop}%
\bibitem [{\citenamefont {Senear}\ and\ \citenamefont
  {Batey}(2002)}]{Senear:2002fm}%
  \BibitemOpen
  \bibfield  {author} {\bibinfo {author} {\bibfnamefont {D.~F.}\ \bibnamefont
  {Senear}}\ and\ \bibinfo {author} {\bibfnamefont {R.}~\bibnamefont {Batey}},\
  }\href@noop {} {\bibfield  {journal} {\bibinfo  {journal} {Biochemistry}\
  }\textbf {\bibinfo {volume} {30}},\ \bibinfo {pages} {6677} (\bibinfo {year}
  {2002})}\BibitemShut {NoStop}%
\bibitem [{\citenamefont {Mascotti}\ and\ \citenamefont
  {Lohman}(1990)}]{Mascotti:1990tg}%
  \BibitemOpen
  \bibfield  {author} {\bibinfo {author} {\bibfnamefont {D.~P.}\ \bibnamefont
  {Mascotti}}\ and\ \bibinfo {author} {\bibfnamefont {T.~M.}\ \bibnamefont
  {Lohman}},\ }\href@noop {} {\bibfield  {journal} {\bibinfo  {journal}
  {Proceedings of the National Academy of Sciences, U.S.A.}\ }\textbf {\bibinfo
  {volume} {87}},\ \bibinfo {pages} {3142} (\bibinfo {year}
  {1990})}\BibitemShut {NoStop}%
\bibitem [{\citenamefont {Mascotti}\ and\ \citenamefont
  {Lohman}(1997)}]{Mascotti:1997th}%
  \BibitemOpen
  \bibfield  {author} {\bibinfo {author} {\bibfnamefont {D.~P.}\ \bibnamefont
  {Mascotti}}\ and\ \bibinfo {author} {\bibfnamefont {T.~M.}\ \bibnamefont
  {Lohman}},\ }\href@noop {} {\bibfield  {journal} {\bibinfo  {journal}
  {Biochemistry}\ }\textbf {\bibinfo {volume} {36}},\ \bibinfo {pages} {7272}
  (\bibinfo {year} {1997})}\BibitemShut {NoStop}%
\bibitem [{\citenamefont {Zhang}\ \emph {et~al.}(1996)\citenamefont {Zhang},
  \citenamefont {Bond}, \citenamefont {Anderson}, \citenamefont {Lohman},\ and\
  \citenamefont {Record}}]{Zhang:1996vn}%
  \BibitemOpen
  \bibfield  {author} {\bibinfo {author} {\bibfnamefont {W.}~\bibnamefont
  {Zhang}}, \bibinfo {author} {\bibfnamefont {J.~P.}\ \bibnamefont {Bond}},
  \bibinfo {author} {\bibfnamefont {C.~F.}\ \bibnamefont {Anderson}}, \bibinfo
  {author} {\bibfnamefont {T.~M.}\ \bibnamefont {Lohman}}, \ and\ \bibinfo
  {author} {\bibfnamefont {M.~T.}\ \bibnamefont {Record}, \bibfnamefont {Jr}},\
  }in\ \href@noop {} {\emph {\bibinfo {booktitle} {Proceedings of the National
  Acedemy of Sciences, U.S.A}}},\ Vol.~\bibinfo {volume} {93}\ (\bibinfo {year}
  {1996})\ p.\ \bibinfo {pages} {2511}\BibitemShut {NoStop}%
\bibitem [{\citenamefont {Datta}\ and\ \citenamefont
  {LiCata}(2003)}]{Datta:2003hl}%
  \BibitemOpen
  \bibfield  {author} {\bibinfo {author} {\bibfnamefont {K.}~\bibnamefont
  {Datta}}\ and\ \bibinfo {author} {\bibfnamefont {V.~J.}\ \bibnamefont
  {LiCata}},\ }\href@noop {} {\bibfield  {journal} {\bibinfo  {journal}
  {Journal of Biological Chemistry}\ }\textbf {\bibinfo {volume} {278}},\
  \bibinfo {pages} {5694} (\bibinfo {year} {2003})}\BibitemShut {NoStop}%
\bibitem [{\citenamefont {Misra}\ \emph
  {et~al.}(1994{\natexlab{a}})\citenamefont {Misra}, \citenamefont {Hecht},
  \citenamefont {Sharp},\ and\ \citenamefont {Friedman}}]{Misra:1994tg}%
  \BibitemOpen
  \bibfield  {author} {\bibinfo {author} {\bibfnamefont {V.~K.}\ \bibnamefont
  {Misra}}, \bibinfo {author} {\bibfnamefont {J.~L.}\ \bibnamefont {Hecht}},
  \bibinfo {author} {\bibfnamefont {K.~A.}\ \bibnamefont {Sharp}}, \ and\
  \bibinfo {author} {\bibfnamefont {R.~A.}\ \bibnamefont {Friedman}},\
  }\href@noop {} {\bibfield  {journal} {\bibinfo  {journal} {Journal of
  molecular Biology}\ }\textbf {\bibinfo {volume} {238}},\ \bibinfo {pages}
  {264} (\bibinfo {year} {1994}{\natexlab{a}})}\BibitemShut {NoStop}%
\bibitem [{\citenamefont {Misra}\ \emph
  {et~al.}(1994{\natexlab{b}})\citenamefont {Misra}, \citenamefont {Sharp},
  \citenamefont {Friedman},\ and\ \citenamefont {Honig}}]{Misra:1994ii}%
  \BibitemOpen
  \bibfield  {author} {\bibinfo {author} {\bibfnamefont {V.~K.}\ \bibnamefont
  {Misra}}, \bibinfo {author} {\bibfnamefont {K.~A.}\ \bibnamefont {Sharp}},
  \bibinfo {author} {\bibfnamefont {R.~A.}\ \bibnamefont {Friedman}}, \ and\
  \bibinfo {author} {\bibfnamefont {B.}~\bibnamefont {Honig}},\ }\href@noop {}
  {\bibfield  {journal} {\bibinfo  {journal} {Journal of molecular biology}\
  }\textbf {\bibinfo {volume} {238}},\ \bibinfo {pages} {245} (\bibinfo {year}
  {1994}{\natexlab{b}})}\BibitemShut {NoStop}%
\bibitem [{\citenamefont {{\AA}berg}\ \emph {et~al.}(2016)\citenamefont
  {{\AA}berg}, \citenamefont {Duderstadt},\ and\ \citenamefont {van
  Oijen}}]{Aberg:2016kc}%
  \BibitemOpen
  \bibfield  {author} {\bibinfo {author} {\bibfnamefont {C.}~\bibnamefont
  {{\AA}berg}}, \bibinfo {author} {\bibfnamefont {K.~E.}\ \bibnamefont
  {Duderstadt}}, \ and\ \bibinfo {author} {\bibfnamefont {A.~M.}\ \bibnamefont
  {van Oijen}},\ }\href@noop {} {\bibfield  {journal} {\bibinfo  {journal}
  {Nucleic Acids Research}\ }\textbf {\bibinfo {volume} {44}},\ \bibinfo
  {pages} {4846} (\bibinfo {year} {2016})}\BibitemShut {NoStop}%
\bibitem [{\citenamefont {Giuntoli}\ \emph {et~al.}(2015)\citenamefont
  {Giuntoli}, \citenamefont {Linzer}, \citenamefont {Banigan}, \citenamefont
  {Sing}, \citenamefont {de~la Cruz}, \citenamefont {Graham}, \citenamefont
  {Johnson},\ and\ \citenamefont {Marko}}]{Giuntoli:2015jf}%
  \BibitemOpen
  \bibfield  {author} {\bibinfo {author} {\bibfnamefont {R.~D.}\ \bibnamefont
  {Giuntoli}}, \bibinfo {author} {\bibfnamefont {N.~B.}\ \bibnamefont
  {Linzer}}, \bibinfo {author} {\bibfnamefont {E.~J.}\ \bibnamefont {Banigan}},
  \bibinfo {author} {\bibfnamefont {C.~E.}\ \bibnamefont {Sing}}, \bibinfo
  {author} {\bibfnamefont {M.~O.}\ \bibnamefont {de~la Cruz}}, \bibinfo
  {author} {\bibfnamefont {J.~S.}\ \bibnamefont {Graham}}, \bibinfo {author}
  {\bibfnamefont {R.~C.}\ \bibnamefont {Johnson}}, \ and\ \bibinfo {author}
  {\bibfnamefont {J.~F.}\ \bibnamefont {Marko}},\ }\href@noop {} {\bibfield
  {journal} {\bibinfo  {journal} {Journal of Molecular Biology}\ }\textbf
  {\bibinfo {volume} {427}},\ \bibinfo {pages} {3123} (\bibinfo {year}
  {2015})}\BibitemShut {NoStop}%
\bibitem [{\citenamefont {Chen}\ \emph {et~al.}(1)\citenamefont {Chen},
  \citenamefont {Santiago}, \citenamefont {Jung}, \citenamefont {ski},
  \citenamefont {Yang}, \citenamefont {Martell}, \citenamefont {Helmann},\ and\
  \citenamefont {Chen}}]{Chen:1jh}%
  \BibitemOpen
  \bibfield  {author} {\bibinfo {author} {\bibfnamefont {T.-Y.}\ \bibnamefont
  {Chen}}, \bibinfo {author} {\bibfnamefont {A.~G.}\ \bibnamefont {Santiago}},
  \bibinfo {author} {\bibfnamefont {W.}~\bibnamefont {Jung}}, \bibinfo {author}
  {\bibfnamefont {L.~u. K.~n.}\ \bibnamefont {ski}}, \bibinfo {author}
  {\bibfnamefont {F.}~\bibnamefont {Yang}}, \bibinfo {author} {\bibfnamefont
  {D.~J.}\ \bibnamefont {Martell}}, \bibinfo {author} {\bibfnamefont {J.~D.}\
  \bibnamefont {Helmann}}, \ and\ \bibinfo {author} {\bibfnamefont
  {P.}~\bibnamefont {Chen}},\ }\href@noop {} {\bibfield  {journal} {\bibinfo
  {journal} {Nature Communications}\ }\textbf {\bibinfo {volume} {6}},\
  \bibinfo {pages} {1} (\bibinfo {year} {1})}\BibitemShut {NoStop}%
\bibitem [{\citenamefont {Graham}\ \emph {et~al.}(2011)\citenamefont {Graham},
  \citenamefont {Johnson},\ and\ \citenamefont {Marko}}]{Graham:2011cy}%
  \BibitemOpen
  \bibfield  {author} {\bibinfo {author} {\bibfnamefont {J.~S.}\ \bibnamefont
  {Graham}}, \bibinfo {author} {\bibfnamefont {R.~C.}\ \bibnamefont {Johnson}},
  \ and\ \bibinfo {author} {\bibfnamefont {J.~F.}\ \bibnamefont {Marko}},\
  }\href@noop {} {\bibfield  {journal} {\bibinfo  {journal} {Nucleic Acids
  Research}\ }\textbf {\bibinfo {volume} {39}},\ \bibinfo {pages} {2249}
  (\bibinfo {year} {2011})}\BibitemShut {NoStop}%
\bibitem [{\citenamefont {Sing}\ \emph {et~al.}(2014)\citenamefont {Sing},
  \citenamefont {Olvera de~la Cruz},\ and\ \citenamefont
  {Marko}}]{Sing:2014dz}%
  \BibitemOpen
  \bibfield  {author} {\bibinfo {author} {\bibfnamefont {C.~E.}\ \bibnamefont
  {Sing}}, \bibinfo {author} {\bibfnamefont {M.}~\bibnamefont {Olvera de~la
  Cruz}}, \ and\ \bibinfo {author} {\bibfnamefont {J.~F.}\ \bibnamefont
  {Marko}},\ }\href@noop {} {\bibfield  {journal} {\bibinfo  {journal} {Nucleic
  Acids Research}\ }\textbf {\bibinfo {volume} {42}},\ \bibinfo {pages} {3783}
  (\bibinfo {year} {2014})}\BibitemShut {NoStop}%
\bibitem [{\citenamefont {Dahlke}\ and\ \citenamefont
  {Sing}(2017)}]{Dahlke:2017bn}%
  \BibitemOpen
  \bibfield  {author} {\bibinfo {author} {\bibfnamefont {K.}~\bibnamefont
  {Dahlke}}\ and\ \bibinfo {author} {\bibfnamefont {C.~E.}\ \bibnamefont
  {Sing}},\ }\href@noop {} {\bibfield  {journal} {\bibinfo  {journal}
  {Biophysical Journal}\ }\textbf {\bibinfo {volume} {112}},\ \bibinfo {pages}
  {543} (\bibinfo {year} {2017})}\BibitemShut {NoStop}%
\bibitem [{\citenamefont {Luo}\ \emph {et~al.}(2014)\citenamefont {Luo},
  \citenamefont {North}, \citenamefont {Rose},\ and\ \citenamefont
  {Poirier}}]{Luo:2014ff}%
  \BibitemOpen
  \bibfield  {author} {\bibinfo {author} {\bibfnamefont {Y.}~\bibnamefont
  {Luo}}, \bibinfo {author} {\bibfnamefont {J.~A.}\ \bibnamefont {North}},
  \bibinfo {author} {\bibfnamefont {S.~D.}\ \bibnamefont {Rose}}, \ and\
  \bibinfo {author} {\bibfnamefont {M.~G.}\ \bibnamefont {Poirier}},\
  }\href@noop {} {\bibfield  {journal} {\bibinfo  {journal} {Nucleic Acids
  Research}\ }\textbf {\bibinfo {volume} {42}},\ \bibinfo {pages} {3017}
  (\bibinfo {year} {2014})}\BibitemShut {NoStop}%
\bibitem [{\citenamefont {Kunzelmann}\ \emph {et~al.}(2010)\citenamefont
  {Kunzelmann}, \citenamefont {Morris}, \citenamefont {Chavda}, \citenamefont
  {Eccleston},\ and\ \citenamefont {Webb}}]{Kunzelmann:2010ie}%
  \BibitemOpen
  \bibfield  {author} {\bibinfo {author} {\bibfnamefont {S.}~\bibnamefont
  {Kunzelmann}}, \bibinfo {author} {\bibfnamefont {C.}~\bibnamefont {Morris}},
  \bibinfo {author} {\bibfnamefont {A.~P.}\ \bibnamefont {Chavda}}, \bibinfo
  {author} {\bibfnamefont {J.~F.}\ \bibnamefont {Eccleston}}, \ and\ \bibinfo
  {author} {\bibfnamefont {M.~R.}\ \bibnamefont {Webb}},\ }\href@noop {}
  {\bibfield  {journal} {\bibinfo  {journal} {Biochemistry}\ }\textbf {\bibinfo
  {volume} {49}},\ \bibinfo {pages} {843} (\bibinfo {year} {2010})}\BibitemShut
  {NoStop}%
\bibitem [{\citenamefont {Kamar}\ \emph {et~al.}(2017)\citenamefont {Kamar},
  \citenamefont {Banigan}, \citenamefont {Erba{\c s}}, \citenamefont
  {Giuntoli}, \citenamefont {Olvera de~la Cruz}, \citenamefont {Johnson},\ and\
  \citenamefont {Marko}}]{Kamar:2017dd}%
  \BibitemOpen
  \bibfield  {author} {\bibinfo {author} {\bibfnamefont {R.~I.}\ \bibnamefont
  {Kamar}}, \bibinfo {author} {\bibfnamefont {E.~J.}\ \bibnamefont {Banigan}},
  \bibinfo {author} {\bibfnamefont {A.}~\bibnamefont {Erba{\c s}}}, \bibinfo
  {author} {\bibfnamefont {R.~D.}\ \bibnamefont {Giuntoli}}, \bibinfo {author}
  {\bibfnamefont {M.}~\bibnamefont {Olvera de~la Cruz}}, \bibinfo {author}
  {\bibfnamefont {R.~C.}\ \bibnamefont {Johnson}}, \ and\ \bibinfo {author}
  {\bibfnamefont {J.~F.}\ \bibnamefont {Marko}},\ }\href@noop {} {\bibfield
  {journal} {\bibinfo  {journal} {Proceedings of the National Academy of
  Sciences, U.S.A}\ }\textbf {\bibinfo {volume} {114}},\ \bibinfo {pages}
  {E3251} (\bibinfo {year} {2017})}\BibitemShut {NoStop}%
\bibitem [{\citenamefont {Gibb}\ \emph {et~al.}(2014)\citenamefont {Gibb},
  \citenamefont {Ye}, \citenamefont {Gergoudis}, \citenamefont {Kwon},
  \citenamefont {Niu}, \citenamefont {Sung},\ and\ \citenamefont
  {Greene}}]{Gibb:2014kf}%
  \BibitemOpen
  \bibfield  {author} {\bibinfo {author} {\bibfnamefont {B.}~\bibnamefont
  {Gibb}}, \bibinfo {author} {\bibfnamefont {L.~F.}\ \bibnamefont {Ye}},
  \bibinfo {author} {\bibfnamefont {S.~C.}\ \bibnamefont {Gergoudis}}, \bibinfo
  {author} {\bibfnamefont {Y.}~\bibnamefont {Kwon}}, \bibinfo {author}
  {\bibfnamefont {H.}~\bibnamefont {Niu}}, \bibinfo {author} {\bibfnamefont
  {P.}~\bibnamefont {Sung}}, \ and\ \bibinfo {author} {\bibfnamefont {E.~C.}\
  \bibnamefont {Greene}},\ }\href@noop {} {\bibfield  {journal} {\bibinfo
  {journal} {PLOS ONE}\ }\textbf {\bibinfo {volume} {9}},\ \bibinfo {pages}
  {e87922} (\bibinfo {year} {2014})}\BibitemShut {NoStop}%
\bibitem [{\citenamefont {Hadizadeh}\ \emph {et~al.}(2016)\citenamefont
  {Hadizadeh}, \citenamefont {Johnson},\ and\ \citenamefont
  {Marko}}]{Hadizadeh:2016hh}%
  \BibitemOpen
  \bibfield  {author} {\bibinfo {author} {\bibfnamefont {N.}~\bibnamefont
  {Hadizadeh}}, \bibinfo {author} {\bibfnamefont {R.~C.}\ \bibnamefont
  {Johnson}}, \ and\ \bibinfo {author} {\bibfnamefont {J.~F.}\ \bibnamefont
  {Marko}},\ }\href@noop {} {\bibfield  {journal} {\bibinfo  {journal} {Journal
  of Bacteriology}\ }\textbf {\bibinfo {volume} {198}},\ \bibinfo {pages}
  {1735} (\bibinfo {year} {2016})}\BibitemShut {NoStop}%
\bibitem [{\citenamefont {Kim}\ \emph {et~al.}(2012)\citenamefont {Kim},
  \citenamefont {Eggel}, \citenamefont {Tarchevskaya}, \citenamefont {Vogel},
  \citenamefont {Prinz},\ and\ \citenamefont {Jardetzky}}]{Kim:2012gq}%
  \BibitemOpen
  \bibfield  {author} {\bibinfo {author} {\bibfnamefont {B.}~\bibnamefont
  {Kim}}, \bibinfo {author} {\bibfnamefont {A.}~\bibnamefont {Eggel}}, \bibinfo
  {author} {\bibfnamefont {S.~S.}\ \bibnamefont {Tarchevskaya}}, \bibinfo
  {author} {\bibfnamefont {M.}~\bibnamefont {Vogel}}, \bibinfo {author}
  {\bibfnamefont {H.}~\bibnamefont {Prinz}}, \ and\ \bibinfo {author}
  {\bibfnamefont {T.~S.}\ \bibnamefont {Jardetzky}},\ }\href@noop {} {\bibfield
   {journal} {\bibinfo  {journal} {Nature}\ }\textbf {\bibinfo {volume}
  {491}},\ \bibinfo {pages} {613} (\bibinfo {year} {2012})}\BibitemShut
  {NoStop}%
\bibitem [{\citenamefont {Pennington}\ \emph {et~al.}(2016)\citenamefont
  {Pennington}, \citenamefont {Tarchevskaya}, \citenamefont {Brigger},
  \citenamefont {Sathiyamoorthy}, \citenamefont {Graham}, \citenamefont
  {Nadeau}, \citenamefont {Eggel},\ and\ \citenamefont
  {Jardetzky}}]{Pennington:2016es}%
  \BibitemOpen
  \bibfield  {author} {\bibinfo {author} {\bibfnamefont {L.~F.}\ \bibnamefont
  {Pennington}}, \bibinfo {author} {\bibfnamefont {S.}~\bibnamefont
  {Tarchevskaya}}, \bibinfo {author} {\bibfnamefont {D.}~\bibnamefont
  {Brigger}}, \bibinfo {author} {\bibfnamefont {K.}~\bibnamefont
  {Sathiyamoorthy}}, \bibinfo {author} {\bibfnamefont {M.~T.}\ \bibnamefont
  {Graham}}, \bibinfo {author} {\bibfnamefont {K.~C.}\ \bibnamefont {Nadeau}},
  \bibinfo {author} {\bibfnamefont {A.}~\bibnamefont {Eggel}}, \ and\ \bibinfo
  {author} {\bibfnamefont {T.~S.}\ \bibnamefont {Jardetzky}},\ }\href@noop {}
  {\bibfield  {journal} {\bibinfo  {journal} {Nature Communications}\ }\textbf
  {\bibinfo {volume} {7}},\ \bibinfo {pages} {1} (\bibinfo {year}
  {2016})}\BibitemShut {NoStop}%
\bibitem [{\citenamefont {Solis}\ and\ \citenamefont {de~la
  Cruz}(2000)}]{Solis:2000dp}%
  \BibitemOpen
  \bibfield  {author} {\bibinfo {author} {\bibfnamefont {F.~J.}\ \bibnamefont
  {Solis}}\ and\ \bibinfo {author} {\bibfnamefont {M.~O.}\ \bibnamefont {de~la
  Cruz}},\ }\href@noop {} {\bibfield  {journal} {\bibinfo  {journal} {European
  Physical Journal E}\ }\textbf {\bibinfo {volume} {4}},\ \bibinfo {pages}
  {143} (\bibinfo {year} {2000})}\BibitemShut {NoStop}%
\bibitem [{\citenamefont {Raspaud}\ \emph {et~al.}(1998)\citenamefont
  {Raspaud}, \citenamefont {de~la Cruz}, \citenamefont {Sikorav},\ and\
  \citenamefont {Livolant}}]{Raspaud:1998hs}%
  \BibitemOpen
  \bibfield  {author} {\bibinfo {author} {\bibfnamefont {E.}~\bibnamefont
  {Raspaud}}, \bibinfo {author} {\bibfnamefont {M.~O.}\ \bibnamefont {de~la
  Cruz}}, \bibinfo {author} {\bibfnamefont {J.~L.}\ \bibnamefont {Sikorav}}, \
  and\ \bibinfo {author} {\bibfnamefont {F.}~\bibnamefont {Livolant}},\
  }\href@noop {} {\bibfield  {journal} {\bibinfo  {journal} {Biophysical
  Journal}\ }\textbf {\bibinfo {volume} {74}},\ \bibinfo {pages} {381}
  (\bibinfo {year} {1998})}\BibitemShut {NoStop}%
\bibitem [{\citenamefont {Tsai}\ \emph {et~al.}(2016)\citenamefont {Tsai},
  \citenamefont {Zhang}, \citenamefont {Zheng},\ and\ \citenamefont
  {Wolynes}}]{Tsai:2016gj}%
  \BibitemOpen
  \bibfield  {author} {\bibinfo {author} {\bibfnamefont {M.-Y.}\ \bibnamefont
  {Tsai}}, \bibinfo {author} {\bibfnamefont {B.}~\bibnamefont {Zhang}},
  \bibinfo {author} {\bibfnamefont {W.}~\bibnamefont {Zheng}}, \ and\ \bibinfo
  {author} {\bibfnamefont {P.~G.}\ \bibnamefont {Wolynes}},\ }\href@noop {}
  {\bibfield  {journal} {\bibinfo  {journal} {Journal of the American Chemical
  Society}\ }\textbf {\bibinfo {volume} {138}},\ \bibinfo {pages} {13497}
  (\bibinfo {year} {2016})}\BibitemShut {NoStop}%
\bibitem [{\citenamefont {Haas}\ \emph {et~al.}(1999)\citenamefont {Haas},
  \citenamefont {Drenth},\ and\ \citenamefont {Wilson}}]{Haas:1999gu}%
  \BibitemOpen
  \bibfield  {author} {\bibinfo {author} {\bibfnamefont {C.}~\bibnamefont
  {Haas}}, \bibinfo {author} {\bibfnamefont {J.}~\bibnamefont {Drenth}}, \ and\
  \bibinfo {author} {\bibfnamefont {W.~W.}\ \bibnamefont {Wilson}},\
  }\href@noop {} {\bibfield  {journal} {\bibinfo  {journal} {The Journal of
  Physical Chemistry B}\ }\textbf {\bibinfo {volume} {103}},\ \bibinfo {pages}
  {2808} (\bibinfo {year} {1999})}\BibitemShut {NoStop}%
\bibitem [{\citenamefont {Oberholzer}\ and\ \citenamefont
  {Lenhoff}(1999)}]{Oberholzer:1999bj}%
  \BibitemOpen
  \bibfield  {author} {\bibinfo {author} {\bibfnamefont {M.~R.}\ \bibnamefont
  {Oberholzer}}\ and\ \bibinfo {author} {\bibfnamefont {A.~M.}\ \bibnamefont
  {Lenhoff}},\ }\href@noop {} {\bibfield  {journal} {\bibinfo  {journal}
  {Langmuir}\ }\textbf {\bibinfo {volume} {15}},\ \bibinfo {pages} {3905}
  (\bibinfo {year} {1999})}\BibitemShut {NoStop}%
\bibitem [{\citenamefont {Stigter}\ and\ \citenamefont
  {Dill}(2002)}]{Stigter:2002dy}%
  \BibitemOpen
  \bibfield  {author} {\bibinfo {author} {\bibfnamefont {D.}~\bibnamefont
  {Stigter}}\ and\ \bibinfo {author} {\bibfnamefont {K.~A.}\ \bibnamefont
  {Dill}},\ }\href@noop {} {\bibfield  {journal} {\bibinfo  {journal}
  {Biochemistry}\ }\textbf {\bibinfo {volume} {29}},\ \bibinfo {pages} {1262}
  (\bibinfo {year} {2002})}\BibitemShut {NoStop}%
\bibitem [{\citenamefont {Brockman}\ \emph {et~al.}(1999)\citenamefont
  {Brockman}, \citenamefont {Frutos},\ and\ \citenamefont
  {Corn}}]{Brockman:1999iu}%
  \BibitemOpen
  \bibfield  {author} {\bibinfo {author} {\bibfnamefont {J.~M.}\ \bibnamefont
  {Brockman}}, \bibinfo {author} {\bibfnamefont {A.~G.}\ \bibnamefont
  {Frutos}}, \ and\ \bibinfo {author} {\bibfnamefont {R.~M.}\ \bibnamefont
  {Corn}},\ }\href@noop {} {\bibfield  {journal} {\bibinfo  {journal} {Journal
  of the American Chemical Society}\ }\textbf {\bibinfo {volume} {121}},\
  \bibinfo {pages} {8044} (\bibinfo {year} {1999})}\BibitemShut {NoStop}%
\bibitem [{\citenamefont {Kremer}\ and\ \citenamefont
  {Grest}(1990)}]{Kremer:1990bn}%
  \BibitemOpen
  \bibfield  {author} {\bibinfo {author} {\bibfnamefont {K.}~\bibnamefont
  {Kremer}}\ and\ \bibinfo {author} {\bibfnamefont {G.~S.}\ \bibnamefont
  {Grest}},\ }\href@noop {} {\bibfield  {journal} {\bibinfo  {journal} {The
  Journal of Chemical Physics}\ }\textbf {\bibinfo {volume} {92}},\ \bibinfo
  {pages} {5057} (\bibinfo {year} {1990})}\BibitemShut {NoStop}%
\bibitem [{\citenamefont {Grest}\ and\ \citenamefont
  {Murat}(1993)}]{Grest:1993uc}%
  \BibitemOpen
  \bibfield  {author} {\bibinfo {author} {\bibfnamefont {G.~S.}\ \bibnamefont
  {Grest}}\ and\ \bibinfo {author} {\bibfnamefont {M.}~\bibnamefont {Murat}},\
  }\href@noop {} {\bibfield  {journal} {\bibinfo  {journal} {Macromolecules}\
  }\textbf {\bibinfo {volume} {26}},\ \bibinfo {pages} {3108} (\bibinfo {year}
  {1993})}\BibitemShut {NoStop}%
\bibitem [{\citenamefont {Plimpton}(1995)}]{Plimpton:1995wla}%
  \BibitemOpen
  \bibfield  {author} {\bibinfo {author} {\bibfnamefont {S.}~\bibnamefont
  {Plimpton}},\ }\href@noop {} {\bibfield  {journal} {\bibinfo  {journal}
  {Journal of Computational Physics}\ }\textbf {\bibinfo {volume} {117}},\
  \bibinfo {pages} {1} (\bibinfo {year} {1995})}\BibitemShut {NoStop}%
\bibitem [{\citenamefont {Manning}(1969)}]{Manning:1969us}%
  \BibitemOpen
  \bibfield  {author} {\bibinfo {author} {\bibfnamefont {G.~S.}\ \bibnamefont
  {Manning}},\ }\href@noop {} {\bibfield  {journal} {\bibinfo  {journal} {The
  Journal of Chemical Physics}\ }\textbf {\bibinfo {volume} {51}},\ \bibinfo
  {pages} {924} (\bibinfo {year} {1969})}\BibitemShut {NoStop}%
\bibitem [{\citenamefont {Humphrey}\ \emph {et~al.}(1996)\citenamefont
  {Humphrey}, \citenamefont {Dalke},\ and\ \citenamefont {Schulten}}]{vmd}%
  \BibitemOpen
  \bibfield  {author} {\bibinfo {author} {\bibfnamefont {W.}~\bibnamefont
  {Humphrey}}, \bibinfo {author} {\bibfnamefont {A.}~\bibnamefont {Dalke}}, \
  and\ \bibinfo {author} {\bibfnamefont {K.}~\bibnamefont {Schulten}},\
  }\href@noop {} {\bibfield  {journal} {\bibinfo  {journal} {Journal of
  molecular graphics}\ }\textbf {\bibinfo {volume} {14}},\ \bibinfo {pages}
  {33} (\bibinfo {year} {1996})}\BibitemShut {NoStop}%
\bibitem [{\citenamefont {Sushko}\ \emph {et~al.}(2016)\citenamefont {Sushko},
  \citenamefont {Thomas}, \citenamefont {Pabit}, \citenamefont {Pollack},
  \citenamefont {Onufriev},\ and\ \citenamefont {Baker}}]{Sushko:2016fy}%
  \BibitemOpen
  \bibfield  {author} {\bibinfo {author} {\bibfnamefont {M.~L.}\ \bibnamefont
  {Sushko}}, \bibinfo {author} {\bibfnamefont {D.~G.}\ \bibnamefont {Thomas}},
  \bibinfo {author} {\bibfnamefont {S.~A.}\ \bibnamefont {Pabit}}, \bibinfo
  {author} {\bibfnamefont {L.}~\bibnamefont {Pollack}}, \bibinfo {author}
  {\bibfnamefont {A.~V.}\ \bibnamefont {Onufriev}}, \ and\ \bibinfo {author}
  {\bibfnamefont {N.~A.}\ \bibnamefont {Baker}},\ }\href@noop {} {\bibfield
  {journal} {\bibinfo  {journal} {Biophysical Journal}\ }\textbf {\bibinfo
  {volume} {110}},\ \bibinfo {pages} {315} (\bibinfo {year}
  {2016})}\BibitemShut {NoStop}%
\bibitem [{\citenamefont {Muthukumar}(1997)}]{Muthukumar:1997tu}%
  \BibitemOpen
  \bibfield  {author} {\bibinfo {author} {\bibfnamefont {M.}~\bibnamefont
  {Muthukumar}},\ }\href@noop {} {\bibfield  {journal} {\bibinfo  {journal}
  {Journal of Chemical Physics}\ }\textbf {\bibinfo {volume} {107}},\ \bibinfo
  {pages} {2619} (\bibinfo {year} {1997})}\BibitemShut {NoStop}%
\bibitem [{\citenamefont {Erba{\c s}}\ and\ \citenamefont
  {Netz}(2013)}]{Erbas:2013ta}%
  \BibitemOpen
  \bibfield  {author} {\bibinfo {author} {\bibfnamefont {A.}~\bibnamefont
  {Erba{\c s}}}\ and\ \bibinfo {author} {\bibfnamefont {R.~R.}\ \bibnamefont
  {Netz}},\ }\href@noop {} {\bibfield  {journal} {\bibinfo  {journal}
  {Biophysical Journal}\ }\textbf {\bibinfo {volume} {104}},\ \bibinfo {pages}
  {1285} (\bibinfo {year} {2013})}\BibitemShut {NoStop}%
\bibitem [{\citenamefont {Tan}\ \emph {et~al.}(2016)\citenamefont {Tan},
  \citenamefont {Terakawa},\ and\ \citenamefont {Takada}}]{Tan:2016ff}%
  \BibitemOpen
  \bibfield  {author} {\bibinfo {author} {\bibfnamefont {C.}~\bibnamefont
  {Tan}}, \bibinfo {author} {\bibfnamefont {T.}~\bibnamefont {Terakawa}}, \
  and\ \bibinfo {author} {\bibfnamefont {S.}~\bibnamefont {Takada}},\
  }\href@noop {} {\bibfield  {journal} {\bibinfo  {journal} {Journal of the
  American Chemical Society}\ }\textbf {\bibinfo {volume} {138}},\ \bibinfo
  {pages} {8512} (\bibinfo {year} {2016})}\BibitemShut {NoStop}%
\bibitem [{\citenamefont {de~la Cruz}\ \emph {et~al.}(1995)\citenamefont {de~la
  Cruz}, \citenamefont {Belloni}, \citenamefont {Delsanti}, \citenamefont
  {Dalbiez}, \citenamefont {Spalla},\ and\ \citenamefont
  {Drifford}}]{delaCruz:1995vz}%
  \BibitemOpen
  \bibfield  {author} {\bibinfo {author} {\bibfnamefont {M.~O.}\ \bibnamefont
  {de~la Cruz}}, \bibinfo {author} {\bibfnamefont {L.}~\bibnamefont {Belloni}},
  \bibinfo {author} {\bibfnamefont {M.}~\bibnamefont {Delsanti}}, \bibinfo
  {author} {\bibfnamefont {J.~P.}\ \bibnamefont {Dalbiez}}, \bibinfo {author}
  {\bibfnamefont {O.}~\bibnamefont {Spalla}}, \ and\ \bibinfo {author}
  {\bibfnamefont {M.}~\bibnamefont {Drifford}},\ }\href@noop {} {\bibfield
  {journal} {\bibinfo  {journal} {The Journal of Chemical Physics}\ }\textbf
  {\bibinfo {volume} {103}},\ \bibinfo {pages} {5781} (\bibinfo {year}
  {1995})}\BibitemShut {NoStop}%
\bibitem [{\citenamefont {Rubinstein}\ and\ \citenamefont
  {Colby}(2003)}]{Rubinstein:2003vd}%
  \BibitemOpen
  \bibfield  {author} {\bibinfo {author} {\bibfnamefont {M.}~\bibnamefont
  {Rubinstein}}\ and\ \bibinfo {author} {\bibfnamefont {R.~H.}\ \bibnamefont
  {Colby}},\ }\href@noop {} {\emph {\bibinfo {title} {{Polymer physics}}}}\
  (\bibinfo  {publisher} {Oxford University Press, USA},\ \bibinfo {year}
  {2003})\BibitemShut {NoStop}%
\bibitem [{\citenamefont {Honig}\ and\ \citenamefont
  {Nicholls}(1995)}]{Honig:1995tt}%
  \BibitemOpen
  \bibfield  {author} {\bibinfo {author} {\bibfnamefont {B.}~\bibnamefont
  {Honig}}\ and\ \bibinfo {author} {\bibfnamefont {A.}~\bibnamefont
  {Nicholls}},\ }\href@noop {} {\bibfield  {journal} {\bibinfo  {journal}
  {Science}\ }\textbf {\bibinfo {volume} {268}},\ \bibinfo {pages} {1144}
  (\bibinfo {year} {1995})}\BibitemShut {NoStop}%
\bibitem [{\citenamefont {Bonthuis}\ \emph {et~al.}(2011)\citenamefont
  {Bonthuis}, \citenamefont {Gekle},\ and\ \citenamefont
  {Netz}}]{Bonthuis:2011cg}%
  \BibitemOpen
  \bibfield  {author} {\bibinfo {author} {\bibfnamefont {D.~J.}\ \bibnamefont
  {Bonthuis}}, \bibinfo {author} {\bibfnamefont {S.}~\bibnamefont {Gekle}}, \
  and\ \bibinfo {author} {\bibfnamefont {R.~R.}\ \bibnamefont {Netz}},\
  }\href@noop {} {\bibfield  {journal} {\bibinfo  {journal} {Physical Review
  Letters}\ }\textbf {\bibinfo {volume} {107}},\ \bibinfo {pages} {508}
  (\bibinfo {year} {2011})}\BibitemShut {NoStop}%
\bibitem [{\citenamefont {Fahrenberger}\ \emph {et~al.}(2015)\citenamefont
  {Fahrenberger}, \citenamefont {Hickey}, \citenamefont {Smiatek},\ and\
  \citenamefont {Holm}}]{Fahrenberger:2015ku}%
  \BibitemOpen
  \bibfield  {author} {\bibinfo {author} {\bibfnamefont {F.}~\bibnamefont
  {Fahrenberger}}, \bibinfo {author} {\bibfnamefont {O.~A.}\ \bibnamefont
  {Hickey}}, \bibinfo {author} {\bibfnamefont {J.}~\bibnamefont {Smiatek}}, \
  and\ \bibinfo {author} {\bibfnamefont {C.}~\bibnamefont {Holm}},\ }\href@noop
  {} {\bibfield  {journal} {\bibinfo  {journal} {Physical Review Letters}\
  }\textbf {\bibinfo {volume} {115}},\ \bibinfo {pages} {118301} (\bibinfo
  {year} {2015})}\BibitemShut {NoStop}%
\bibitem [{\citenamefont {Li}\ \emph {et~al.}(2016)\citenamefont {Li},
  \citenamefont {Erba{\c s}}, \citenamefont {Zwanikken},\ and\ \citenamefont
  {Olvera de~la Cruz}}]{Li:2016cu}%
  \BibitemOpen
  \bibfield  {author} {\bibinfo {author} {\bibfnamefont {H.}~\bibnamefont
  {Li}}, \bibinfo {author} {\bibfnamefont {A.}~\bibnamefont {Erba{\c s}}},
  \bibinfo {author} {\bibfnamefont {J.}~\bibnamefont {Zwanikken}}, \ and\
  \bibinfo {author} {\bibfnamefont {M.}~\bibnamefont {Olvera de~la Cruz}},\
  }\href@noop {} {\bibfield  {journal} {\bibinfo  {journal} {Macromolecules}\
  }\textbf {\bibinfo {volume} {49}},\ \bibinfo {pages} {9239} (\bibinfo {year}
  {2016})}\BibitemShut {NoStop}%
\bibitem [{\citenamefont {Parsaeian}\ \emph {et~al.}(2013)\citenamefont
  {Parsaeian}, \citenamefont {de~la Cruz},\ and\ \citenamefont
  {Marko}}]{Parsaeian:2013iu}%
  \BibitemOpen
  \bibfield  {author} {\bibinfo {author} {\bibfnamefont {A.}~\bibnamefont
  {Parsaeian}}, \bibinfo {author} {\bibfnamefont {M.~O.}\ \bibnamefont {de~la
  Cruz}}, \ and\ \bibinfo {author} {\bibfnamefont {J.~F.}\ \bibnamefont
  {Marko}},\ }\href@noop {} {\bibfield  {journal} {\bibinfo  {journal}
  {Physical Review E}\ }\textbf {\bibinfo {volume} {88}},\ \bibinfo {pages}
  {40703} (\bibinfo {year} {2013})}\BibitemShut {NoStop}%
\bibitem [{\citenamefont {Knotts}\ \emph {et~al.}(2007)\citenamefont {Knotts},
  \citenamefont {Rathore}, \citenamefont {Schwartz},\ and\ \citenamefont
  {de~Pablo}}]{KnottsIV:2007gn}%
  \BibitemOpen
  \bibfield  {author} {\bibinfo {author} {\bibfnamefont {T.~A.}\ \bibnamefont
  {Knotts}, \bibfnamefont {IV}}, \bibinfo {author} {\bibfnamefont
  {N.}~\bibnamefont {Rathore}}, \bibinfo {author} {\bibfnamefont {D.~C.}\
  \bibnamefont {Schwartz}}, \ and\ \bibinfo {author} {\bibfnamefont {J.~J.}\
  \bibnamefont {de~Pablo}},\ }\href@noop {} {\bibfield  {journal} {\bibinfo
  {journal} {The Journal of Chemical Physics}\ }\textbf {\bibinfo {volume}
  {126}},\ \bibinfo {pages} {084901} (\bibinfo {year} {2007})}\BibitemShut
  {NoStop}%
\bibitem [{\citenamefont {Uusitalo}\ \emph {et~al.}(2015)\citenamefont
  {Uusitalo}, \citenamefont {Ing{\'o}lfsson}, \citenamefont {Akhshi},
  \citenamefont {Tieleman},\ and\ \citenamefont {Marrink}}]{Uusitalo:2015eh}%
  \BibitemOpen
  \bibfield  {author} {\bibinfo {author} {\bibfnamefont {J.~J.}\ \bibnamefont
  {Uusitalo}}, \bibinfo {author} {\bibfnamefont {H.~I.}\ \bibnamefont
  {Ing{\'o}lfsson}}, \bibinfo {author} {\bibfnamefont {P.}~\bibnamefont
  {Akhshi}}, \bibinfo {author} {\bibfnamefont {D.~P.}\ \bibnamefont
  {Tieleman}}, \ and\ \bibinfo {author} {\bibfnamefont {S.~J.}\ \bibnamefont
  {Marrink}},\ }\href@noop {} {\bibfield  {journal} {\bibinfo  {journal}
  {Journal of Chemical Theory and Computation}\ }\textbf {\bibinfo {volume}
  {11}},\ \bibinfo {pages} {3932} (\bibinfo {year} {2015})}\BibitemShut
  {NoStop}%
\end{thebibliography}%

\end{document}